\documentclass[12pt]{article}

\usepackage{pifont}

\usepackage{dcolumn}
\newcolumntype{d}{D{.}{.}{-1} }

\usepackage{latexsym}
\usepackage{amsthm}
\usepackage{amsmath,amssymb,amsfonts,bm,amsbsy,bbm}
\usepackage{epsfig}
\usepackage{graphicx,rotating,lscape}
\usepackage{verbatim}
\usepackage{multirow,color}
\usepackage{geometry}
\usepackage{setspace}
\usepackage{subfigure}
\usepackage{bbm}
\usepackage{ulem} 
\newcommand{\stkout}[1]{\ifmmode\text{\sout{\ensuremath{#1}}}\else\sout{#1}\fi}

\usepackage{enumerate}
\usepackage{url} 
\usepackage{tikz, adjustbox, float} 
\usepackage{dsfont} 
\usepackage{mathrsfs} 
\usepackage{soul} 
\usepackage{bm} 
\usepackage{booktabs}  
\usepackage{enumitem} 

\usepackage[ruled,vlined]{algorithm2e}
\SetKwInOut{Input}{Input}
\SetKwInOut{Output}{Output}
\SetKwRepeat{Do}{do}{while}%

\usepackage{graphicx}
\usepackage{color}
\usepackage{nicefrac}
\usepackage{float}
\usepackage{caption}

\DeclareMathOperator*{\simiid}{{\overset{\text{iid}}{\sim}}}
\DeclareMathOperator*{\todist}{{\overset{\text{d}}{\to}}}
\DeclareMathOperator*{\toprob}{{\overset{\text{p}}{\to}}}
\DeclareMathOperator*{\eqdist}{{\overset{\text{d}}{=}}}

\def\real{{\mathbb{R}}}
\def\integer{\mathbb{N}}

\DeclareMathOperator*{\cov}{cov}

\def\CI{\text{CI}}

\def\0{{\bf 0}}
\def\1{{\bf 1}}

\def\btheta{{\bm{\theta}}}
\def\htheta{\widehat{\theta}}
\def\hbtheta{\widehat{\bm{\theta}}}
\def\hbSigma{\widehat{\bm{\Sigma}}}
\def\bSigma{\bm{\Sigma}}
\def\hsigma{\widehat{\sigma}}

\def\bpi{{\bm{\pi}}}

\def\bdeta{{\bm{\eta}}}
\def\hbdeta{\widehat{{\bm{\eta}}}}

\def\bOmega{{\bm{\Omega}}}

\DeclareMathOperator*{\var}{var}

\DeclareMathOperator*{\argzero}{argzero}

\def\mux{{\mu_x}}
\def\muy{{\mu_y}}
\def\nx{{n_x}}
\def\ny{{n_y}}
\def\barx{{\overline{X}}}
\def\bary{{\overline{Y}}}
\def\sigmax{{\sigma_x}}
\def\sigmay{{\sigma_y}}

\def\pix{{\pi_x}}

\def\hsigmax{{\widehat{\sigma}_x}}
\def\hsigmay{{\widehat{\sigma}_y}}
\def\halpha{{\widehat{\alpha}}}

\def\wh{\widehat}
\def\wt{\widetilde}

\def\var{\hbox{var}}
\def\cov{\hbox{cov}}

\def\bar{\overline}

\def\bse{\begin{eqnarray*}}
\def\ese{\end{eqnarray*}}
\def\be{\begin{eqnarray}}
\def\ee{\end{eqnarray}}
\def\bsq{\begin{equation*}}
\def\esq{\end{equation*}}
\def\bq{\begin{equation}}
\def\eq{\end{equation}}

\usepackage{natbib}
\setcitestyle{citesep={,}}
\setcitestyle{aysep={}}

\usepackage[linktoc=all,colorlinks=true,linkcolor=black,citecolor=blue,urlcolor=blue,bookmarks=false,hypertexnames=true]{hyperref}

\def\boxit#1{\vbox{\hrule\hbox{\vrule\kern6pt  \vbox{\kern6pt#1\kern6pt}\kern6pt\vrule}\hrule}}

\begin{document}

\def\spacingset#1{\renewcommand{\baselinestretch}%
{#1}\small\normalsize} \spacingset{1}

\begin{center}
    {\LARGE\bf Bridging the gap between experimental burden and statistical power for quantiles equivalence testing}
\end{center}

\vskip 0.5cm
\begin{center}
$\textsc{Jun Wu}^{1}$, 
\textsc{St\'ephane Guerrier}$^{2\text{-}4}$, 
$\textsc{Si Gou}^{2,3}$, \\
$\textsc{Yogeshvar N. Kalia}^{2,3}$ 
\& $\textsc{Luca Insolia}^{2\text{-}4}$ 
\end{center}

\begin{center}
 \footnotesize
    $^{1}$Geneva School of Economics and Management, University of Geneva, Switzerland;
    $^{2}$School of Pharmaceutical Sciences, University of Geneva, Switzerland;
    $^{3}$Institute of Pharmaceutical Sciences of Western Switzerland, University of Geneva, Switzerland;
    $^{4}$Department of Earth Sciences, University of Geneva, Switzerland.
\end{center}  

\bigskip
\begin{abstract}
\noindent
Testing the equivalence of multiple quantiles between two populations is important in many scientific applications, such as clinical trials, where conventional mean-based methods may be inadequate. This is particularly relevant in bridging studies that compare drug responses across different experimental conditions or patient populations. These studies often aim to assess whether a proposed dose for a target population achieves pharmacokinetic levels comparable to those of a reference population where efficacy and safety have been established. The focus is on extreme quantiles which directly inform both efficacy and safety assessments. When analyzing heterogeneous Gaussian samples, where a single quantile of interest is estimated, the existing Two One-Sided Tests method for quantile equivalence testing (qTOST) tends to be overly conservative. To mitigate this behavior, we introduce $\alpha$-qTOST, a finite-sample adjustment that achieves uniformly higher power compared to qTOST while maintaining the test size at the nominal level. Moreover, we extend the quantile equivalence framework to simultaneously assess equivalence across multiple quantiles. Through theoretical guarantees and an extensive simulation study, we demonstrate that $\alpha$-qTOST offers substantial improvements, especially when testing extreme quantiles under heteroskedasticity and with small, unbalanced sample sizes. We illustrate these advantages through two case studies, one in HIV drug development, where a bridging clinical trial examines exposure distributions between male and female populations with unbalanced sample sizes, and another in assessing the reproducibility of an identical experimental protocol performed by different operators for generating biodistribution profiles of topically administered and locally acting products.
\end{abstract}

\noindent
{\it Keywords:} 
bioequivalence, bridging studies, finite-sample adjustment, heterogeneous populations, two one-sided tests.

\spacingset{1.9} 

\section{Introduction}

Equivalence testing determines whether an effect of interest is sufficiently similar across two populations \citep{wellek2010testing}.
It is critical in pharmaceutical research, where it is commonly denoted as BioEquivalence (BE), particularly when comparing formulations (or drug products), doses, or treatments across different populations (see e.g.,~\citealt{patterson2017bioequivalence}). 
Unlike traditional hypothesis testing that aims to detect differences, BE aims to establish that two formulations or treatments are sufficiently similar within predetermined equivalence margins.
BE studies play a key role in the approval of generic drugs and in the assessment of post-market modifications, by comparing pharmacokinetic (PK) parameters of two formulations administered to healthy subjects, typically in randomized crossover designs \citep{chow1999design,du2015likelihood}. 
These evaluations focus on key PK metrics such as the area under the concentration-time curve (AUC), maximum concentration (C$_{\text{max}}$), and time to maximum concentration (T$_{\text{max}}$), which serve as surrogate measures for drug absorption extent and rate.
Traditionally, regulatory agencies have predominantly relied on average BE assessments, which compare the mean value of PK parameters between formulations \citep{us2001fda, committee2010ema, berger1996bioequivalence}.
However, average BE may be inadequate when features of the drug exposure distribution other than the central tendency are of interest. For instance, a formulation could demonstrate acceptable average BE while still producing markedly different responses in a certain proportion of individuals, potentially compromising therapeutic outcomes.
These issues commonly arise in the presence of heteroskedasticity, unbalanced sample sizes, or clinically significant extreme values, where alternative approaches may be necessary to ensure accurate BE assessments.
This limitation has prompted the development of more comprehensive BE approaches, such as population and individual BE (see e.g.,~\citealt{anderson1990consideration, schall1993population, gould2000practical, chow2003statistical}). 
While these approaches for BE assessments represent important advances, they may not adequately address specific regulatory concerns in certain therapeutic contexts. 
For example, a more targeted approach may be necessary for drugs with narrow therapeutic indices (i.e.,~where the margin between effective and toxic doses is small), or for conditions where low drug concentrations could lead to treatment failure or the development of drug resistance. 

In this work, we focus on the approach for quantile BE developed by \cite{pei2008statistical} to compare a given quantile between two normal populations.
This offers a more comprehensive framework for comparing drug exposure between heterogeneous populations, and it can specifically address concerns about both efficacy and safety at critical thresholds associated with more extreme quantiles. For instance, in the context of systemic drugs, comparing lower quantiles of the PK distribution may indicate a higher risk of treatment failure or development of drug resistance for one population, while a comparison of upper quantiles could signal potential toxicity of the drug \citep{benet1995bioequivalence,endrenyi2013determination,yu2015novel}.
Understanding the impact of low PK levels could be applied to the management of 
persistent viral infections such as HIV, where the rapid replication and mutation rates of viruses frequently result in resistant variants (see e.g.,~\citealp{pillay1998antiviral, little2002antiretroviral, wu2005modeling, nascimento2020pharmacokinetic}). These variants emerge under suboptimal drug suppression, undermining treatment efficacy and increasing the risk of viral transmission within subjects (see e.g.,~ \citealp{monforte1998clinical}, \citealp{huang2003modeling}, \citealp{oette2006predictors}, \citealp{nair2014pharmacokinetics}, \citealp{gopalan2017sub}, \citealp{soeria2024sub}). 
Moreover, quantile BE is particularly relevant in bridging studies, which aim to leverage existing clinical data from one well-studied reference population to support drug approval in a target population \citep{liu2004bridging,chow2012statistical}. These studies are typical in scenarios where conducting full clinical trials on both populations would be impractical or unethical, such as in pediatric drug development \citep{ich2001pediatric} or multi-regional trials \citep{chow2010bridging}, where ensuring comparable pharmacological responses between populations at some critical quantiles is crucial.

While \cite{pei2008statistical} introduced a \textit{Two One-Sided Tests} (TOST; \citealp{schuirmann1987comparison}) procedure for quantile equivalence testing (qTOST in short), this procedure can be overly conservative in finite samples.
To mitigate this issue, we propose $\alpha$-qTOST, a simple adjustment that leads to a uniformly more powerful test than the existing qTOST, while maintaining the test size at the nominal level $\alpha$. 
Moreover, we extend the existing quantile equivalence testing framework to the simultaneous assessment of multiple quantiles.
We establish the theoretical properties of $\alpha$-qTOST and demonstrate its advantages through an extensive simulation study and two case studies.
The first case study is related to HIV pharmacotherapy, where antiretroviral efficacy and safety profiles may vary significantly across patient populations (see e.g.,~\citealp{leth2006pharmacokinetic,daskapan2019systematic,calcagno2021impact,toledo2023pharmacokinetics}). This is a cause of concern in patients exhibiting low PK parameters, corresponding to low quantiles of the PK distribution, who may fail to achieve therapeutic drug concentrations necessary for viral suppression, potentially resulting in treatment failure or the development of drug resistance (see e.g.,~\citealp{orrell2016effect,monforte1998clinical,oette2006predictors,wu2005modeling,nascimento2020pharmacokinetic,huang2003modeling,nair2014pharmacokinetics}). 
The second case study is related to the topical administration of locally acting therapeutic agents for the treatment of dermatologic conditions (although it is equally applicable for cosmeceutical ingredients) and the development of methodologies to determine the spatial distribution of the compounds \citep{quartier2019cutaneous} and their use for assessments of equivalence. As a first step, it is necessary to demonstrate that the method can be reproduced between different operators. Therefore, here we describe assessments of equivalence at two quantiles of interest, for an identical experimental protocol performed by different operators in the context of topical products.
Importantly, the proposed $\alpha$-qTOST is applicable not only to clinical trials and pre-clinical data, but also to a variety of other contexts where equivalence testing is used, such as psychology \citep{Lakens:18}, engineering \citep{moore2022engineering}, software development \citep{dolado2014equivalence}, and social sciences \citep{aggarwal20232}. 

\subsection{Organization and notation}

The article is organized as follows. Section~\ref{sec:framework} presents the existing framework for quantile BE and the resulting qTOST procedure when testing a single quantile. Section~\ref{sec:pi_tost} introduces the proposed $\alpha$-qTOST adjustment, as well as its statistical properties and algorithmic implementation. Section~\ref{sec:pitost_mvt} extends the quantile equivalence testing framework to the simultaneous assessment of multiple quantiles, and generalizes the qTOST and $\alpha$-qTOST procedures to such cases. Section~\ref{sec:simul} compares the finite-sample performances of the two approaches through an extensive simulation study, both when testing a single quantile and jointly assessing two quantiles. Section~\ref{sec:case_study} illustrates the advantages of the proposed approach through two illustrative bridging studies. 
Finally, in Section~\ref{sec:final} we provide some final remarks and directions for future research.

We complete this section by defining the notation used throughout the paper. We denote with $\Phi$ and $\phi$ the cumulative and the density distribution function of a standard normal random variable, respectively, and indicate with $z_\alpha$ the corresponding upper $\alpha$ quantile such that $\Phi(z_{\alpha})=1-\alpha$. 
Moreover, we use standard asymptotic notation.  For sequences of random variables, $\todist$ denotes convergence in distribution and  $\toprob$ denotes convergence in probability. For deterministic positive sequences $\{a_n\}$ and $\{b_n\}$, we write $a_n=O(b_n)$ to indicate that there exists a positive constant $C$ such that $a_n \leq C b_n$ for all sufficiently large $n$, while $a_n=o(b_n)$ indicates that $\lim_{n \to \infty} a_n/b_n = 0$. We write $a_n \asymp  b_n$ to indicate that the sequences are of the same order, meaning that $a_n=O(b_n)$ and $b_n=O(a_n)$. For their stochastic counterparts based on random variable sequences $\{X_n\}$ and $\{Y_n\}$, we write $X_n=O_p(Y_n)$ to indicate that the sequence $X_n/Y_n$ is bounded in probability. 
We also write $X_n=o_p(Y_n)$ to express that $X_n/Y_n \toprob 0$, while $X_n \asymp_p Y_n$ indicates that $X_n = O_p(Y_n)$ and $Y_n = O_p(X_n)$.
Finally, we denote equality in distribution with $\eqdist$.

\section{Quantile equivalence testing}
\label{sec:framework}

In this section, we present the statistical framework for quantile equivalence testing proposed by \cite{pei2008statistical} and the resulting qTOST procedure when assessing a single quantile.
Let $X$ and $Y$ denote two continuous random variables representing the estimated endpoint of interest (e.g.,~a transformation of some PK parameter) across two populations.
For example, $X$ may represent measurements from a reference population where efficacy and safety have been established (e.g.,~adult male patients from Phase III trials), while $Y$ represents measurements from a target population under assessment (e.g.,~patients from different ethnic backgrounds, female patients, or pediatric patients). 
For these reasons, the two samples are typically heterogeneous and have unbalanced sample sizes. 

Let $q_x$ be the (unknown) quantile of interest for the reference population, and let $\pi_x \equiv \Pr(X \leq q_x) $, for a given $\pi_x$. 
To assess the treatment effects in the target population, we examine $\pi_y \equiv \Pr(Y \leq q_x)$, which represents the proportion of subjects in the target population with measurements that are below the quantile $q_x$ of the reference population.
To demonstrate quantile equivalence between the two populations, we test whether $\pi_y$ is sufficiently close to $\pi_x$. Therefore, the following hypotheses are considered: 
\be\label{eq:hyp_pi}
    \text{H}_0: \; \pi_y \not\in \Pi_1
    \quad \text{vs.} \quad
    \text{H}_1: \; \pi_y \in \Pi_1 \equiv  (\pi_x + \Delta_l, \pi_x + \Delta_u) , 
\ee
where $(\Delta_l, \Delta_u)$ represent some fixed equivalence margins (i.e.,~they cannot be random). For instance, these margins can be based on expert domain knowledge or regulatory guidance.
Although asymmetric margins may be more appropriate in some applications (e.g.,~when considering very extreme quantiles), without much loss of generality, we restrict our attention to the conventional choice of symmetric equivalence margins around $\pi_x$ by taking $c \equiv \Delta_u=-\Delta_l$.

To test the hypotheses in \eqref{eq:hyp_pi}, \cite{pei2008statistical} proposed a TOST-like procedure under Gaussian assumptions. 
Namely, consider two samples 
\be \label{eq:canonical}
X_i \simiid \mathcal{N}(\mu_x, \sigma_x^2) \quad \text{and} \quad Y_j \simiid \mathcal{N}(\mu_y, \sigma_y^2) ,
\ee
where $i = 1, \ldots, \nx$ and $j = 1, \ldots, \ny$, 
and $\cov\left(X_{i}, Y_{j}\right) = 0$ for all $i,j$'s.
The normality assumption is a reasonable approximation in many practical applications, such as bridging clinical trials, where the two samples often represent measurements of the (transformation of) PK responses taken from the two groups \citep{julious2000pharmacokinetic,wellek2010testing,patterson2017bioequivalence}.
From the definition of $\pi_x$, we have
\be \label{eq:quantiles}
\begin{aligned} 
\pi_x &= \Pr\left(X_{i} \leq q_x\right) = \Phi\left(\frac{q_x - \mu_x}{\sigma_x}\right) \iff q_x = \mu_x + \sigma_x \Phi^{-1}(\pi_x),   \\
\pi_y &= \Pr\left(Y_{j} \leq q_x\right) = \Phi\left(\frac{q_x - \mu_y}{\sigma_y}\right) = \Phi(\theta), 
\end{aligned}
\ee
and
\be \label{eq:theta}
\theta \equiv \frac{\mu_x - \mu_y}{\sigma_y} + \frac{\sigma_x}{\sigma_y}\Phi^{-1}(\pi_x).
\ee
Therefore, the hypotheses in \eqref{eq:hyp_pi} can be equivalently formulated as
\be\label{eq:hyp_theta}
    \text{H}_0: \; \theta \not\in \Theta_1
    \quad \text{vs.} \quad
    \text{H}_1: \; \theta \in \Theta_1  \equiv  \left( \delta_l , \delta_u \right) , 
\ee
where $\delta_l \equiv \Phi^{-1}(\pi_x +\Delta_l)$ and $ \delta_u \equiv \Phi^{-1}(\pi_x + \Delta_u)$.
Then, a natural plug-in estimator for $\theta$ is given by
\be \label{eq:mle_theta}
\htheta \equiv \frac{\bar{X} - \bar{Y}}{\hsigma_y} + \frac{\hsigma_x}{\hsigma_y}\Phi^{-1}(\pi_{x}), 
\ee
where 
$$
\bar{X} \equiv \frac{1}{n_x} \sum_{{i = 1}}^{n_x} X_{i} \sim \mathcal{N}(\mu_x,\sigma_x^2/n_x)  \quad \text{and} \quad  \bar{Y} \equiv \frac{1}{n_y} \sum_{{j = 1}}^{n_y} Y_{j} \sim \mathcal{N}(\mu_y,\sigma_y^2/n_y),
$$
and 
\be \label{eq:mle_sigmas}
\hsigma_x^2 \equiv \frac{1}{\nu_x} \sum_{{i = 1}}^{n_x} (X_i - \bar{X})^2 \eqdist \frac{\sigma_x^2}{\nu_x} W_x  \quad \text{and} \quad  \hsigma_y^2 \equiv \frac{1}{\nu_y} \sum_{{j = 1}}^{n_y} (Y_j - \bar{Y})^2 \eqdist \frac{\sigma_y^2}{\nu_y} W_y, 
\ee
with $\nu_x\equiv\nx-1$ and $\nu_y\equiv\ny-1$, where
$W_x \sim \chi^2_{\nu_x}$ is independent of $W_y \sim \chi^2_{\nu_y}$, and
$\chi^2_{\nu}$ denotes a chi-square distribution with $\nu$ degrees of freedom.
\cite{pei2008statistical} showed that, as $n_y \to \infty$ with $l \equiv n_y/n_x$ held constant, $\htheta$ in \eqref{eq:mle_theta} satisfies
\be \label{eq:theta_hat_asym_dist}
\sqrt{n_y}(\htheta - \theta) \todist \mathcal{N}\left(0, \sigma_a^2 \right), \;\; \text{where} \;\; \sigma_a^2 = 1 + \frac{\theta^2}{2} + \frac{l}{\gamma}\left[1 + \frac{\{\Phi^{-1}(\pi_x)\}^2}{2}\right] \;\; \text{and} \;\; \gamma \equiv \frac{\sigma_y^2}{\sigma_x^2} .
\ee
Based on $\wh\sigma^2 \equiv \wh\sigma_a^2 / n_y$, with $\wh\sigma_a^2$ being an estimator of $\sigma_a^2$ based on the plug-in estimates $\htheta$ and $\wh \gamma$, one can construct an asymptotic $100(1-2\alpha)\%$ confidence interval for $\theta$ as
\be \label{eq:ci_tost}
\CI_\alpha \equiv ( \htheta - z_{\alpha} \wh\sigma , \htheta + z_{\alpha} \wh\sigma ) .
\ee
Therefore, based on the Interval-Inclusion Principle (IIP; \citealt{berger1996bioequivalence}), the corresponding TOST-like procedure leads to a declaration of BE if $\CI_\alpha \subset \Theta_1$ in \eqref{eq:hyp_theta}.

\section{The $\alpha$-qTOST adjusted procedure}
\label{sec:pi_tost}

Based on the confidence interval in \eqref{eq:ci_tost}, the rejection region of the qTOST procedure can be expressed as
\be \label{eq:tost_rej_region}
    R(\alpha) \equiv \left\{\wh\theta \in \real,
      \wh\sigma \in \real_{>0} :\,
      z_\alpha \wh\sigma + \delta_l < \htheta< \delta_u -z_\alpha \wh\sigma \right\}.
\ee
Therefore, its probability of rejecting H$_0$ is given by 
\be\label{eq:tost_power}
    \omega(\theta, \sigma, \alpha) \equiv
    \Pr \left( \CI_{\alpha} \subset \Theta_1 \; \big\lvert \; \theta, \sigma, \alpha \right) = 
    \Pr \left\{ (\htheta, \hsigma)^T \in R(\alpha) \; \big\lvert \; \theta, \sigma, \alpha \right\}
    ,
\ee 
where we ignore the dependency on $n_x, n_y, \delta_l$ and $\delta_u$ to simplify the notation, as these are fixed known quantities.
The size of the qTOST procedure is defined as the supremum of the probability of rejecting H$_0$ in \eqref{eq:tost_power} over the space of the null hypothesis (see e.g.,~\citealp{Lehmann:1986}), that is,
$
\sup_{\theta \not\in \Theta_1} \, \omega(\theta, \sigma, \alpha) 
$.
The theoretical results presented in Section~\ref{sec:theory} suggest that the qTOST is quite conservative, in the sense that $\sup_{\theta \not\in \Theta_1} \, \omega(\theta, \sigma, \alpha)  < \alpha$, and in many settings that are of practical interest can be considerably smaller than $\alpha$. This, in turn, may lead to a substantial loss in statistical power for the qTOST procedure (i.e.,~a reduction in the probability of rejecting H$_0$ when $\theta \in \Theta_1$). This is also illustrated by the simulation results presented in Section~\ref{sec:simul_univ}, which highlight that such a conservative behavior is more pronounced in settings of great scientific interest, such as heterogeneous populations with uneven sample sizes.

To mitigate the conservativeness of qTOST, we develop a finite-sample adjustment by matching its size to the nominal level $\alpha$, thereby increasing the test power.
We denote the resulting procedure as $\alpha$-qTOST, which can be viewed as an extension of the $\alpha$-TOST \citep{boulaguiem2024finite}, tailored for average equivalence testing problems, to the context of quantile equivalence.
Specifically, $\alpha$-qTOST replaces the nominal significance level $\alpha$ employed in \eqref{eq:tost_rej_region} with an adjusted level $\alpha^*$ defined as 
\begin{equation} \label{eq:alpha_star}
\alpha^* \equiv \alpha^*(\sigma) = \argzero_{\xi \in [\alpha,0.5)} \left\{ \sup_{\theta \not\in \Theta_1} \, \omega(\theta, \sigma, \xi)  -\alpha \right\} ,
\end{equation}
where we omit for simplicity the dependency of $\alpha^*(\sigma)$ on fixed known quantities. 
Computational details on how we solve the matching problem in \eqref{eq:alpha_star} are presented in Section~\ref{sec:computational}.
Importantly, when $\alpha^*$ in \eqref{eq:alpha_star} exists, the $\alpha$-qTOST procedure ensures a (theoretical) size of $\alpha$, and our results presented in Section~\ref{sec:theory} suggest that $\alpha^*$ exists and is unique under very mild conditions. Moreover, since $\alpha^* \geq \alpha$, the $\alpha$-qTOST procedure provides shorter confidence intervals than the qTOST and therefore leads to a uniformly more powerful test procedure.
In particular, similarly to \eqref{eq:ci_tost}, we construct confidence intervals 
$
\CI_{\alpha^*} \equiv ( \htheta - z_{\alpha^*} \wh\sigma , \htheta + z_{\alpha^*} \wh\sigma ) ,
$
and based on the IIP reject H$_0$ in \eqref{eq:hyp_theta} when $\CI_{\alpha^*} \in \Theta_1$.

We remark that $\alpha^*$ in \eqref{eq:alpha_star} is a theoretical adjustment that depends on the unknown $\sigma$.
Therefore, a natural estimator for $\alpha^*$ in \eqref{eq:alpha_star} depending on $\hsigma$ is given by
\begin{equation} \label{eq:alpha_star_hat}
    \halpha^* \equiv \alpha^*(\hsigma) = \argzero_{\xi \in [\alpha,0.5)} \left\{  \sup_{\theta \not\in \Theta_1} \omega(\theta, \hsigma, \xi)-\alpha \right\} ,
\end{equation}
which can be solved similarly to \eqref{eq:alpha_star}, as described in Section~\ref{sec:computational}.
Notably, as suggested by the theory presented in Section~\ref{sec:theory} and also illustrated by our simulation results in Section~\ref{sec:simul_univ}, the $\alpha$-qTOST procedure based on $\halpha^*$ in \eqref{eq:alpha_star_hat} maintains a level $\alpha$ in finite samples, in the sense that its empirical size does not exceed $\alpha$.
However, the test procedure may become slightly conservative when considering more extreme quantiles $\pi_x$ and smaller, unbalanced sample sizes, while still remaining less conservative than qTOST. While this behavior aligns with the considered asymptotic framework, it is noteworthy that $\alpha$-qTOST does not lead to a liberal test procedure, which is an essential requirement in BE studies as it relates to the ``consumer'' risk \citep{patterson2017bioequivalence}.
Moreover, since $\halpha^* \geq \alpha$ to achieve a size of $\alpha$  in \eqref{eq:alpha_star_hat}, also the $\alpha$-qTOST based on $\halpha^*$ leads to shorter confidence intervals compared to qTOST. Therefore, based on the IIP, it leads to a procedure uniformly more powerful than the existing qTOST. 
Although the results presented in Section~\ref{sec:theory} indicate that the two procedures are asymptotically equivalent, in many realistic scenarios, such as the case study presented in  Section~\ref{sec:case_study_1} characterized by heterogeneous and unbalanced sample sizes, $\alpha$-qTOST leads to a BE declaration when the qTOST fails. This is further supported by our simulation study presented in Section~\ref{sec:simul_univ}, illustrating that $\alpha$-qTOST remains uniformly more powerful than qTOST, leading to power gains close to 30\% in some settings, while effectively controlling the size at the nominal level $\alpha$.

\subsection{Theoretical results}
\label{sec:theory}

Our proposal aims to improve the finite-sample properties of the qTOST based on \eqref{eq:ci_tost}. However, the absence of a closed-form expression for $\omega(\theta, \sigma, \alpha)$ in \eqref{eq:alpha_star} renders it difficult to study the exact finite-sample properties of the qTOST and $\alpha$-qTOST procedures.
Therefore, we adopt the following strategy. First, we consider the case where $\sigma_x$ and $\sigma_y$ in \eqref{eq:theta} are known, and then demonstrate that the difference between the resulting estimator and $\htheta$ in \eqref{eq:mle_theta} is a higher-order term. This result suggests that the properties obtained when $\sigma_x$ and $\sigma_y$ are known allow us to understand the finite-sample properties of the qTOST and $\alpha$-qTOST procedures.
Specifically, throughout this section, we study a simplified estimator for $\theta$ defined as
\be \label{eq:theta_tilde}
    \wt\theta \equiv \frac{\bar{X} - \bar{Y}}{\sigma_y} + \frac{\sigma_x}{\sigma_y}\Phi^{-1}(\pi_{x}) \sim \mathcal{N}(\theta, \tau^2), 
\ee
where
\begin{align*}
    \mathbb{E}\left[\wt\theta \right] &= \frac{\mu_x - \mu_y}{\sigma_y} + \frac{\sigma_x}{\sigma_y}\Phi^{-1}(\pi_{x}) = \theta ,  \\
    \tau^2 &\equiv \var \left( \wt\theta \right) = \frac{1}{\sigma_{y}^2} \left(\frac{\sigma_{x}^2}{n_x} + \frac{\sigma_{y}^2}{n_y}\right) = \frac{n_y \sigma_{x}^2 + n_x \sigma_{y}^2}{\sigma_{y}^2 n_x n_y} . 
\end{align*}
In this case, we can express the probability of rejecting H$_0$ as 
\be\label{eq:tost_power_known_sigmas}
    \wt\omega(\theta, \tau, \alpha) \equiv  
    \Pr\left (z_\alpha \tau + \delta_l <\wt\theta< \delta_u -z_\alpha \tau \; \big\lvert \; \theta, \tau, \alpha \right) .
\ee
Letting $n \equiv \min(n_x, n_y)$, in Appendix~\ref{app:omega_tilde_closeness} we show that 
\be \label{eq:goodnes_omega_tilde}
    \htheta - \theta = O_p(n^{-1/2}) , \quad
    \wt\theta - \theta = O_p(n^{-1/2}) , \quad \text{and } \quad 
    \htheta - \wt\theta = O_p(n^{-1}) . 
\ee
This result suggests the following decomposition 
\bse
    \htheta - \theta = \underbrace{(\htheta - \wt\theta )}_{O_p(n^{-1})} + \underbrace{ (\wt\theta - \theta ) }_{O_p(n^{-1/2})}
    \Longleftrightarrow \htheta - \theta \asymp_p \wt\theta - \theta ,
\ese
where \eqref{eq:goodnes_omega_tilde} ensures that the difference $(\htheta - \wt\theta)$ does not dominate, converging to zero at the rate $1/n$, while $(\wt\theta - \theta)$ converges to zero at the rate $1/\sqrt{n}$. Thus, to characterize the theoretical properties of the qTOST and $\alpha$-qTOST procedures, we restrict our attention to the probability of rejecting H$_0$ based on $\wt\theta$ as presented in \eqref{eq:tost_power_known_sigmas},  which allows us to understand the properties of the considered procedures based on \eqref{eq:tost_power}.

Regarding the conservativeness of qTOST, in Appendix~\ref{app:TOST_conservativeness} we show that
\be \label{eq:tost_size}
\sup_{\theta \not\in \Theta_1} \, \wt\omega(\theta, \tau, \alpha) = \max \{ \wt\omega(\delta_l, \tau, \alpha ), \; \wt\omega(\delta_u, \tau, \alpha ) \} < \alpha ,
\ee
indicating that the qTOST is level-$\alpha$  in finite samples, and only asymptotically achieves size-$\alpha$, in the sense that
$\lim_{\tau \rightarrow 0} \left\{ \sup_{\theta \not\in \Theta_1} \, \wt\omega(\theta, \tau, \alpha ) \right\} = \alpha.
$
For the $\alpha$-qTOST procedure, the counterpart of \eqref{eq:alpha_star} is 
\begin{equation} \label{eq:alpha_star_tilde}
\wt\alpha^* \equiv \wt\alpha^*(\tau) = \argzero_{\xi \in [\alpha,0.5)} \left\{ \sup_{\theta \not\in \Theta_1} \, \wt\omega(\theta, \tau, \xi )  -\alpha \right\} .
\end{equation}
In Appendix~\ref{app:alpha_star_existance}, we establish a necessary and sufficient condition for the existence of $\wt\alpha^*$ in \eqref{eq:alpha_star_tilde}. 
This condition requires that 
$
\tau < (\delta_u - \delta_l)/ \{ \Phi^{-1}(1/2 + \alpha) \} ,
$
which ensures sufficient statistical power to solve the matching problem in \eqref{eq:alpha_star_tilde}. This represents a very mild requirement for operational purposes, which can be verified empirically on given data. 

\subsection{Computational details}
\label{sec:computational}

The $\alpha$-qTOST adjustment in \eqref{eq:alpha_star} can be computed very efficiently. 
To approximate the rejection probability $\omega(\theta, \sigma, \alpha)$ in \eqref{eq:tost_power} in finite samples, we rely on Monte Carlo integration. 
Specifically, $\htheta$ in \eqref{eq:mle_theta} satisfies
$$
\htheta \eqdist a_2 \frac{(a_3 + a_4 Z )}{W_2} + \frac{W_1}{W_2} \frac{a_1}{a_2} \Phi^{-1}(\pi_{x}) , 
$$
where $Z\sim\mathcal{N}(0,1)$, $W_1\sim \chi_{\nu_x}$, $W_2 \sim \chi_{\nu_y}$ are independent and $\chi_{\nu}$ denotes a chi distribution with $\nu$ degrees of freedom,
for fixed $a_1 \equiv \sigmax / \sqrt{\nu_x}$, $a_2 \equiv \sigmay/\sqrt{\nu_y}$, $a_3 \equiv \mux - \muy$, and $a_4 \equiv \sqrt{\sigma_x^2/\nx+\sigma_y^2/\ny}$. This can also be expressed as
$$
\htheta \eqdist b_1\frac{1}{W_2} + b_2 \frac{Z}{W_2} + b_3 \frac{W_1}{W_2}, 
$$
for fixed $b_1 \equiv \{\theta\gamma -\Phi^{-1}(\pi_x)\}\sqrt{\nu_y/\gamma}$, $b_2 \equiv \sqrt{(1/\nx+\gamma/\ny)\nu_y/\gamma}$, and $b_3 \equiv \Phi^{-1}(\pi_x)\sqrt{\nu_y/\nu_x\gamma}$. 
Moreover, we have $\wh\gamma \eqdist    \gamma (W_2^2 \nu_x)/ (W_1^2 \nu_y)$. 
Therefore, for given $\theta$, 
$\gamma$, $\alpha$, $\delta_l$, $\delta_u$, $\nx$ and $\ny$, 
the Monte Carlo procedure to approximate \eqref{eq:tost_power} requires only the generation of realizations for the $Z$, $W_1$ and $W_2$ random variables.
While this approach increases the computational burden compared to the asymptotic counterpart of \eqref{eq:tost_power} based on \eqref{eq:tost_power_known_sigmas}, the additional computational overhead remains limited and does not significantly impact the overall processing time.
Then, we construct $\alpha^*$ in \eqref{eq:alpha_star} using an iterative algorithm where at each iteration $k \in \integer$, we compute
\be \label{eq:alpha_star_algo}
\alpha^{(k+1)} = \alpha + \alpha^{(k)} - 
\omega(\theta, \sigma, \alpha^{(k)}) ,
\ee 
and the algorithm is initialized at  $\alpha^{(0)} = \alpha$. When $|\alpha^{(k+1)} - \alpha^{(k)}|$ is sufficiently small, the iterative algorithm is stopped and provides $\alpha^* = \alpha^{(k)}$.
Moreover, the same algorithm can be used to obtain $\halpha^*$ in \eqref{eq:alpha_star_hat}.
In Appendix~\ref{app:algo}, considering $\wt\omega(\theta, \tau, \alpha^{(k)})$ in place of $\omega(\theta, \sigma, \alpha^{(k)} )$ in \eqref{eq:alpha_star_algo}, we show that this iterative algorithm converges exponentially fast to the target $\wt\alpha^*$.
Namely, there exists some constant $b>0$ such that
$$
\Big\lvert \wt\alpha^{*(k+1)} - \wt\alpha^* \Big\rvert < \frac{1}{2} e^{-bk} .
$$

Finally, the proposed $\alpha$-qTOST is available on the \texttt{cTOST} package in \texttt{R}, which is also available on the GitHub repository
\url{https://github.com/stephaneguerrier/cTOST}.

\section{Extension to multiple quantiles}
\label{sec:pitost_mvt}

In this section, we extend the quantile equivalence testing framework and the resulting $\alpha$-qTOST procedure to joint assessments across a fixed number $K>1$ of quantiles.
Namely, let
$\bpi_x \equiv [\pi_{x_1}, \ldots, \pi_{x_K}]^T$ and $\bpi_y \equiv [\pi_{y_1}, \ldots, \pi_{y_K}]^T$ where, based on \eqref{eq:quantiles}, we consider 
$
\pi_{x_k} \equiv \Pr\left(X_{i} \leq q_{x_k} \right) 
$ 
and 
$
\pi_{y_k} \equiv \Pr\left(Y_{j} \leq q_{x_k} \right)
$,
for $k=1, \ldots, K$.
Considering without loss of generality equivalence margins $\bm{\Delta_l} \equiv \1_K \Delta_l$ and $\bm{\Delta_u} \equiv \1_K \Delta_u$, with $\1_K$ denoting a vector of ones of length $K$, we are thus interested in assessing the hypotheses
\begin{equation}\label{eq:hyp_pi_mvt}
    \text{H}_{0}: \; \bm{\pi_y} \not\in \mathbf{\Pi_1} \quad \text{vs.} \quad \text{H}_{1}: \; \bm{\pi_y} \in \mathbf{\Pi_1} \equiv \bigcap_{k=1}^K  (\pi_{x_k} + \Delta_l, \pi_{x_k} + \Delta_u) .
\end{equation}
As in Section~\ref{sec:framework},
letting $\btheta \equiv [\theta_1, \ldots, \theta_K]^T$, with $\theta_k \equiv \Phi^{-1}(\pi_{y_k})$ for $k=1, \ldots, K$, 
the hypotheses in \eqref{eq:hyp_pi_mvt} can be equivalently formulated as
\be\label{eq:hyp_theta_mvt}
    \text{H}_0: \; \btheta \not\in \bm{\Theta}_1
    \quad \text{vs.} \quad
    \text{H}_1: \; \btheta \in \bm{\Theta}_1  , 
\ee
where 
$\bm{\Theta}_{1} \equiv \{\bm{x} \in\real^K\,\big|\, \delta_{l_k} < x_k < \delta_{u_k} ,\, k=1,\dots,K\} $ defines the $K$-dimensional parallelotope delimited by the equivalence margins $\delta_{l_k} \equiv \Phi^{-1}\{\pi_{x_k} + \Delta_l \}$ and $\delta_{u_k} \equiv \Phi^{-1}\{\pi_{x_k} + \Delta_u \}$.
Following \eqref{eq:mle_theta}, for $k=1, \ldots, K$, we express the estimator for $\theta_k$ as
\be\label{eq:theta_hat_k} 
    \htheta_k = \frac{\bar{X} - \bar{Y}}{\hsigma_y} + \frac{\hsigma_x}{\hsigma_y}\Phi^{-1}(\pi_{x_k}) , 
\ee
and construct $\hbtheta \equiv [\htheta_1, \ldots, \htheta_K]^T$. 
Assuming that $n_y / n_x \asymp 1$ with $ n \asymp n_y \asymp n_x$ such that $n_x / n \to c_1$ and $n_y / n \to c_2$, in Appendix~\ref{app:mvt} we demonstrate that 
\be \label{eq:asymp_theta_dist_mvt}
\sqrt{n}(\hbtheta - \btheta) \todist \mathcal{N}\left(\0, \bSigma_a \right) ,
\ee
and obtain an estimator for $\bSigma \equiv \bSigma_a / n$ which is denoted as $\wh\bSigma$.
Namely, for $j \neq k$, we have 
$$
 \wh\Sigma_{j,k} \equiv \wh\cov(\htheta_j , \htheta_k) = \frac{1}{n} \left[ 1 + \frac{\htheta_j \htheta_k}{2}  +
 \frac{l}{\wh\gamma} \left\{ 1 + \frac{ \Phi^{-1}(\pi_{x_j}) \Phi^{-1}(\pi_{x_k})}{2} 
 \right\} \right] ,
$$
which extends the result in \eqref{eq:theta_hat_asym_dist} to the joint assessment of $K>1$ target quantiles.

An extension of the qTOST procedure in \eqref{eq:ci_tost} to assess equivalence for more than one quantile can be obtained along the lines of the multivariate TOST for average BE \citep{pallmann2017simultaneous}. In particular, equivalence for \eqref{eq:hyp_theta_mvt} is declared only if marginal equivalence is achieved at all $K$ quantiles, and since all tests are asymptotically level $\alpha$, also their intersection leads to a level-$\alpha$ test \citep{berger1996bioequivalence}.
The resulting multiple quantiles TOST, which we still refer to as qTOST for simplicity, extends the approach from Section~\ref{sec:framework} to $K>1$.
Namely, letting $\hsigma_k^2 \equiv \wh\Sigma_{k,k} $, for $k=1,\ldots,K$, it considers the intersection of marginal, asymptotic $100(1-2\alpha)\%$ confidence intervals for $\btheta$:
\be \label{eq:ci_tost_mvt}
\CI_{K, \alpha} \equiv \bigcap_{k=1}^K
\left\{ \htheta_k - z_{\alpha} \wh\sigma_k , \htheta_k + z_{\alpha} \wh\sigma_k \right\} ,
\ee
and it leads to a declaration of BE in \eqref{eq:hyp_theta_mvt} if $\CI_{K,\alpha} \subset \bm{\Theta}_1$.
However, despite its simplicity, the qTOST becomes increasingly conservative as $K$ increases. 

To mitigate the conservativeness of qTOST for quantiles equivalence, we propose a strategy similar to the one developed in Section~\ref{sec:pi_tost} to test a single quantile, and we denote the resulting procedure as $\alpha$-qTOST even when $K>1$.
This approach is similar in spirit to the multivariate $\alpha$-TOST procedure for multivariate average BE \citep{boulaguiem2025multivariate}. 
For the qTOST based on \eqref{eq:ci_tost_mvt}, its probability of rejecting H$_0$ in \eqref{eq:hyp_theta_mvt} can be expressed as 
$$
\omega_K (\btheta, \bSigma, \alpha ) \equiv \Pr (\CI_{K,\alpha} \subset \bm{\Theta}_1) = \Pr \left ( \bigcap_{k=1}^K
\left\{ z_{\alpha} \wh\sigma_k + \delta_{l_k} < \htheta_k < \delta_{u_k} - z_{\alpha} \wh\sigma_k  \right\} \right) .
$$
Thus, the corresponding test size is
$
\sup_{\btheta \not\in \bm{\Theta}_1} \, \omega_K (\btheta, \bSigma, \alpha ) ,
$
and we consider an $\alpha$-qTOST procedure that depends on an adjusted level $\alpha^*$ which is defined as 
\be \label{eq:pi_tost_mvt}
    \alpha^* \equiv \alpha^*(\bSigma) = \argzero_{\xi \in [\alpha,0.5)} \left\{ \sup_{\btheta \not\in \bm{\Theta}_1} \, \omega_K (\btheta, \bSigma, \xi ) -\alpha \right\} .
\ee
Therefore, the $\alpha$-qTOST leads to a BE declaration if $\CI_{K,\alpha^*} \subset \bm{\Theta}_1$.
Similarly to its single-quantile counterpart, we consider the feasible adjustment 
$\halpha^* \equiv \alpha^*(\hbSigma)$, which can be constructed as described in Section~\ref{sec:computational}.

\section{Simulation study}
\label{sec:simul}

This section presents the results of an extensive simulation study comparing the operating characteristics of the qTOST and $\alpha$-qTOST procedures for both single and multiple quantile assessments.

\subsection{Testing a single quantile}
\label{sec:simul_univ}

To assess the hypotheses in \eqref{eq:hyp_pi}, we consider $\alpha=5\%$ and $c=\Delta_u=-\Delta_l>0$.
Based on the distributional assumptions in \eqref{eq:canonical}, without loss of generality, we set $\mu_x = 0$ and $\sigma_x = 1$ for the reference population.
For the target population, we consider 50 equally-spaced values of $\theta$ in \eqref{eq:theta} spanning $[ \Phi^{-1}(\pi_x - 1.2 c),  \Phi^{-1}(\pi_x + 1.2 c) ]$ to characterize the range of $\pi_y$ values. We then consider various reference quantile levels $\pi_x$, variance ratios $\gamma = \sigma_y^2/\sigma_x^2$, and sample size ratios $l = n_y/n_x$.

In the following, we present simulation results for increasing sample sizes for the target group $n_y \in \{ 30, 60, 90, 120 \}$, and $l \in \{ 1, \nicefrac{1}{2}, \nicefrac{1}{3} \}$, representing both balanced ($l=1$) and unbalanced ($l<1$) designs.
As customary in bridging clinical trials, we focus on settings with $n_x \geq n_y$.
To evaluate performance under both homoscedastic and heteroscedastic conditions, we examined variance ratios $\gamma \in \{1, 2\}$.
We consider reference quantile levels $\pi_x \in \{0.05, 0.1, 0.25, 0.5\}$ with corresponding equivalence margins, symmetric around $\pi_x$, of $c = \{0.025, 0.05, 0.1, 0.2\}$, respectively.
We assess performance by examining the probability of rejecting H$_0$ across varying $\theta$ values, allowing us to evaluate the test size, either at $ \theta =  \Phi^{-1}(\pi_x - c)$ or $ \theta =  \Phi^{-1}(\pi_x + c)$ according to \eqref{eq:tost_size}, and the test power for $ \Phi^{-1}(\pi_x - c) < \theta < \Phi^{-1}(\pi_x + c$).
The simulation study is based on $5 \times 10^4$ Monte Carlo replications. 

\begin{figure}[t!]
    \centering
    \hspace*{-1.75cm}
    \includegraphics[width=1.2\textwidth]{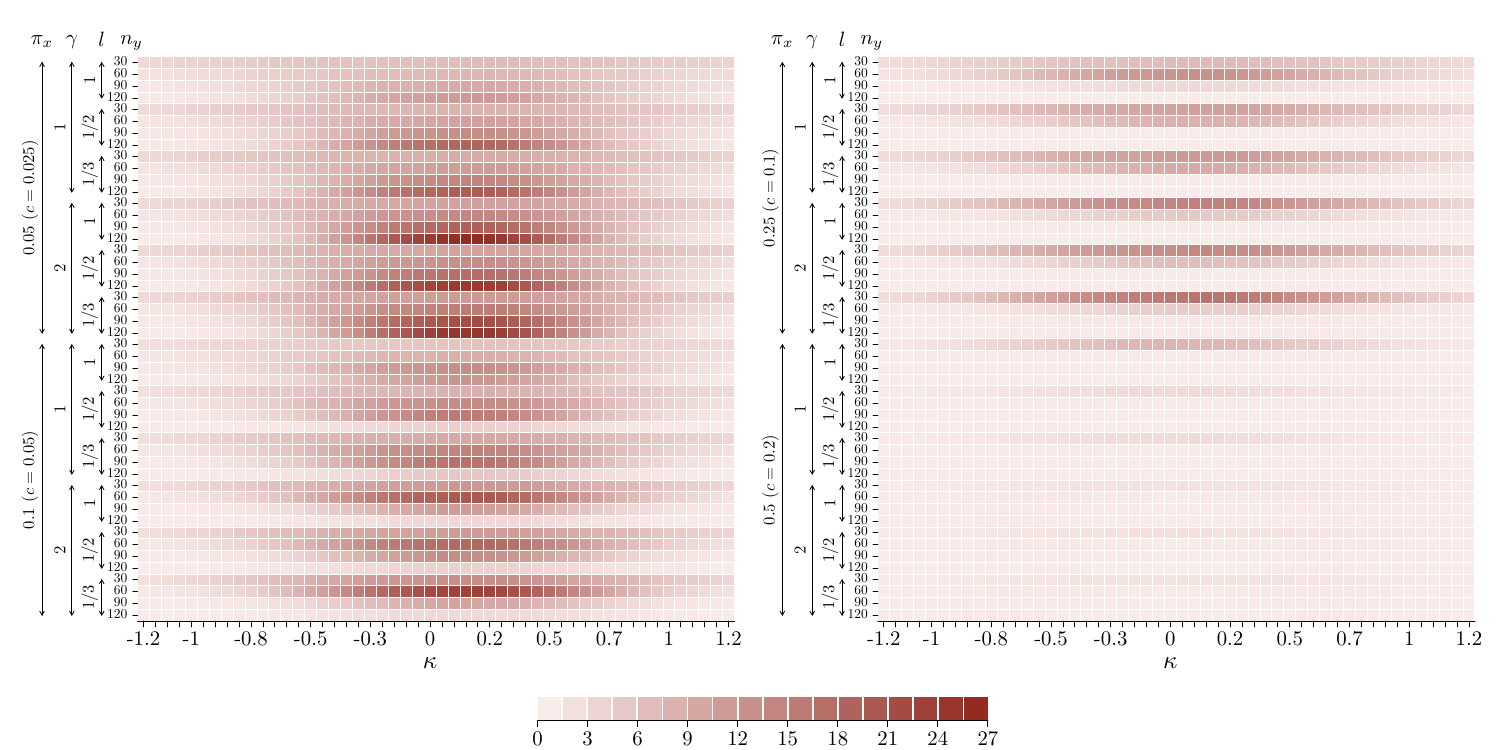}
    \caption{Simulation results comparing the difference in the probability of rejecting H$_0$ at $\theta = \Phi^{-1}(\pi_x + \kappa c) $ for $\alpha$-qTOST with respect to qTOST when varying $n_y$, $l$, $\gamma$, $\pi_x$ and $c$.}
    \label{fig:sim_gain}
\end{figure}
\begin{figure}[t]
    \centering
    \includegraphics[width=1\textwidth]{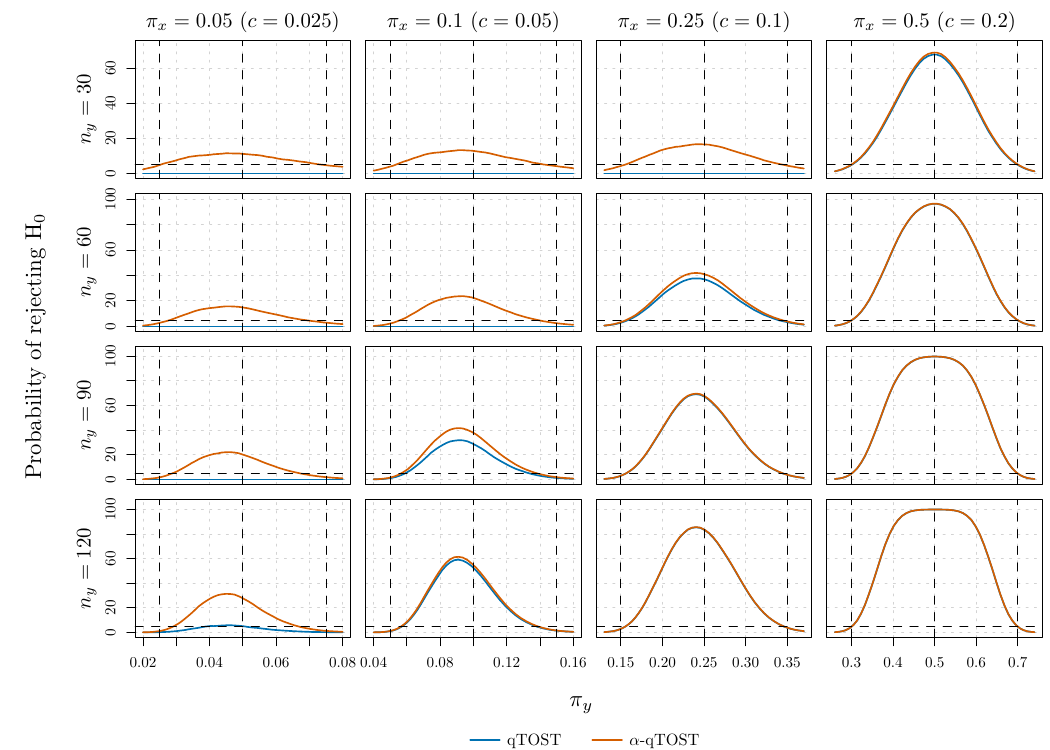}
    \caption{Simulation results comparing the probability of rejecting H$_0$ for $\alpha$-qTOST and qTOST when $\gamma=2$ and $l=\nicefrac{1}{3}$.}
    \label{fig:sim_subset}
\end{figure}

Figure~\ref{fig:sim_gain} illustrates the difference in power of $\alpha$-qTOST relative to qTOST across all simulation scenarios.
To simplify the comparison, we let 
$\theta = \Phi^{-1}(\pi_x + \kappa c) $ and report $\kappa \in [-1.2, 1.2]$
on the x-axis.
Since $\halpha^* \geq \alpha$, $\alpha$-qTOST demonstrates uniformly greater power than qTOST across all settings.
This advantage is more pronounced in challenging scenarios where qTOST reaches very limited power (see also Figure~\ref{fig:sim_subset}).
These include heteroskedastic settings ($\gamma>1$) and unbalanced sample sizes ($l<1$) when testing more extreme quantiles ($\pi_x < 0.25$), yielding power gains of 15-30\% for the $\alpha$-qTOST.
These findings are especially relevant for bridging clinical trials, where these data features are often encountered.
In settings where qTOST achieves an adequate test size, the adjusted level $\halpha^*$ remains close to the nominal level $\alpha$, resulting in comparable performance between the two procedures.
This is further highlighted in Figure~\ref{fig:sim_subset}, which compares the probability of rejecting H$_0$ for the two procedures in the setting with $\gamma=2$ and $l=\nicefrac{1}{3}$. 
Results for other simulation settings show similar patterns and are provided in Appendix~\ref{app:simul}.

\begin{figure}[t]
    \centering
    \includegraphics[width=1\textwidth]{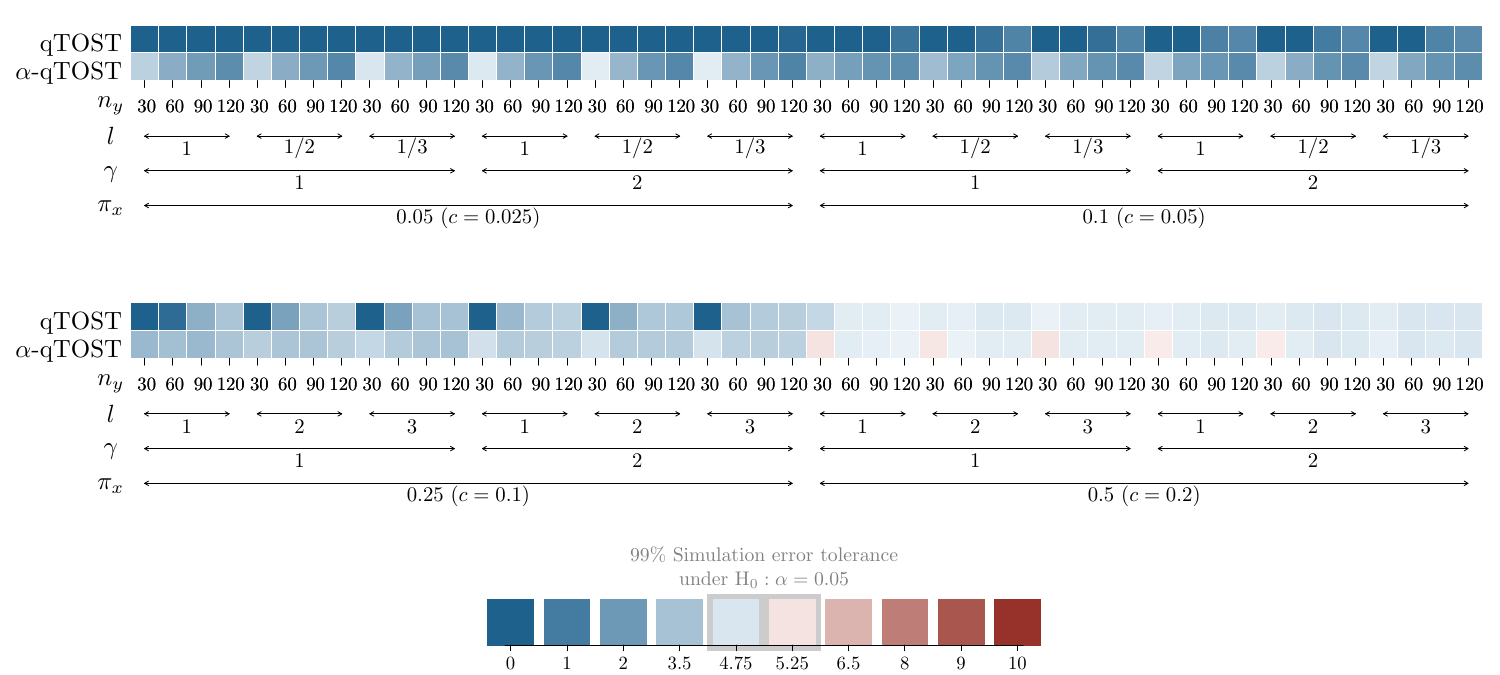}
    \caption{Simulation results comparing the empirical size of qTOST and $\alpha$-qTOST procedures when varying $n_y$, $l$, $\gamma$, $\pi_x$ and $c$.}
    \label{fig:sim_size}
\end{figure}

Regarding empirical size control, Figure~\ref{fig:sim_size} reveals that while qTOST often leads to a very conservative test procedure, particularly in complex scenarios, $\alpha$-qTOST maintains closer adherence to the nominal level $\alpha$, showing only slight conservativeness in more challenging settings. This superior control of the type I error, combined with its substantial gains in power, strongly supports the use of $\alpha$-qTOST over the traditional qTOST, especially in practical applications involving extreme quantiles and/or heteroskedastic settings with small and unbalanced sample sizes.

\subsection{Simultaneous testing of multiple quantiles}
\label{sec:simul_mvt}

In this section, we compare the operating characteristics of qTOST and $\alpha$-qTOST when jointly assessing equivalence on multiple quantiles. 
Motivated by the case study presented in Section~\ref{sec:case_study_2}, we restrict our attention to the hypotheses in \eqref{eq:hyp_pi_mvt} when $K=2$ (i.e.,~considering only two quantiles).
In particular, we take $\pi_{x_1} = 0.2$, $\pi_{x_2} = 0.8$, and $c=\Delta_u=-\Delta_l=0.15$.
This choice allows us to assess quantile equivalence between the tails of the two distributions. 
Moreover, we set $\gamma = 1$, $l=1$, and vary $n_y \in \{ 10, 30, 50\}$.
We present below simulation results for $n_y=30$; results for other values of $n_y$ provide similar conclusions and are presented in Appendix~\ref{app:simul_mvt}.
The nominal significance level is fixed at $\alpha = 0.05$, and the simulation study is based on $5 \times 10^4$ Monte Carlo replications. 
We evaluate performance by comparing the probability of rejecting H$_0$ across varying $\btheta$ values. Namely, similarly to the approach described in Section~\ref{sec:simul_univ}, to characterize the range of $\bpi_y$ values we consider 50 equally-spaced values for $\theta_k$ in the range $[ \Phi^{-1}(\pi_{x_k} - 1.2 c),  \Phi^{-1}(\pi_{x_k} + 1.2 c) ]$, where $k=1,2$.
Based on \eqref{eq:hyp_theta_mvt}, such a grid of $\btheta$ values allows us to assess the test power and level when $\btheta \in \bm{\Theta}_1$ and $\btheta \notin \bm{\Theta}_1$, respectively. 

\begin{figure}[t!]
    \centering
    \includegraphics[width=1\textwidth]{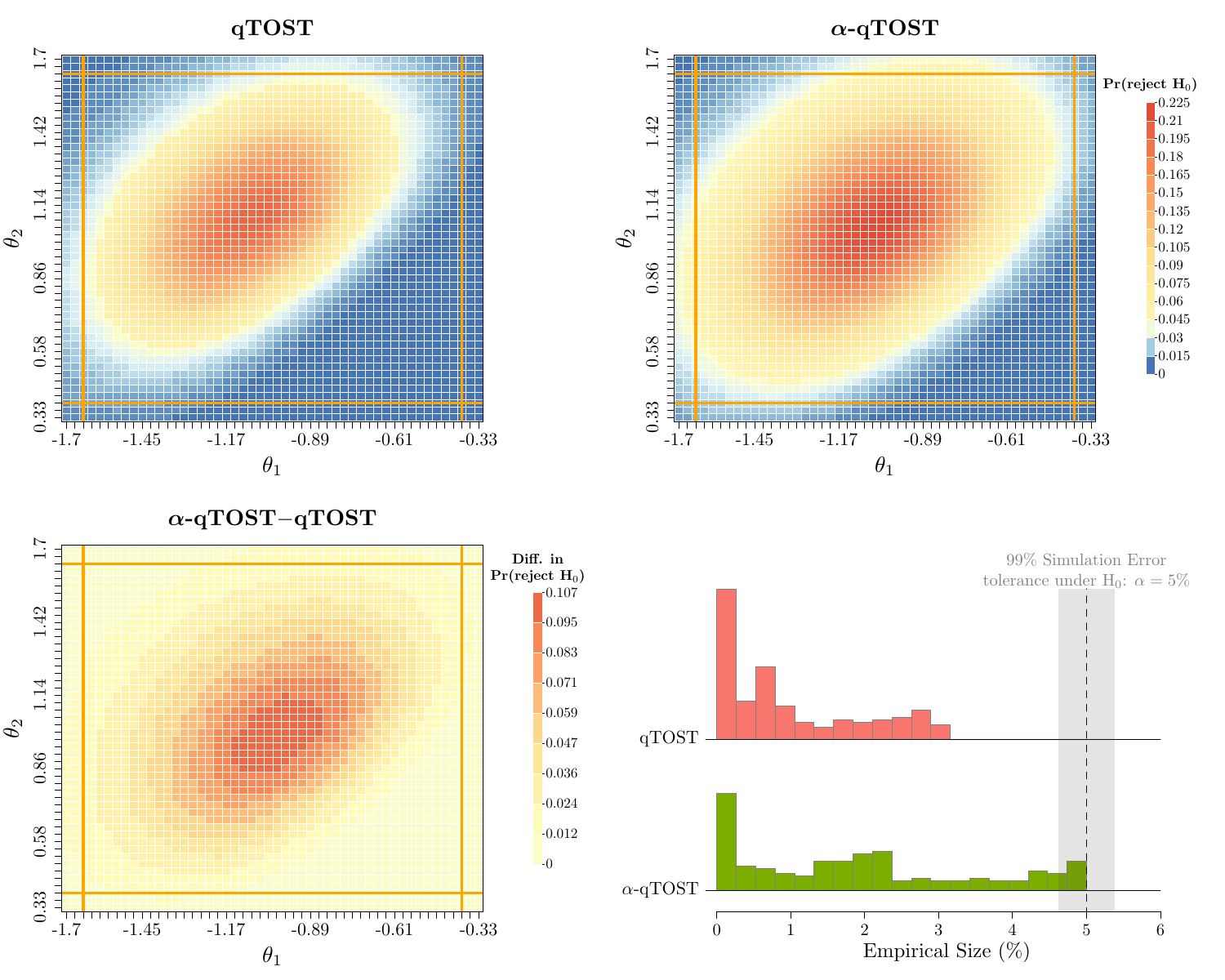}
    \caption{Simulation results comparing the operating characteristics of the qTOST and $\alpha$-qTOST procedures for $n_y = 30$.
    The heatmaps represent the probability of rejecting H$_0$ for the qTOST (top left) and $\alpha$-qTOST (top right) procedures across a grid of $\btheta$ values, as well as the difference between these probabilities (bottom left). For each method, the probability of rejecting H$_0$ along $\btheta$ values that lie on the boundary of the hypothesis space is also reported (bottom right).}
    \label{fig:sim_mvt_1}
\end{figure}
Figure~\ref{fig:sim_mvt_1} compares the probability of rejecting H$_0$ for the qTOST (top left panel) and $\alpha$-qTOST (top right panel) across a grid of $\btheta$ values, as well as the difference between these probabilities (bottom left panel). The thick solid vertical and horizontal lines highlighted in orange represent the equivalence margins associated to each $\theta_k$ parameter. Similarly to the single quantile case presented in Section \ref{sec:simul_univ}, these heatmaps illustrate that the $\alpha$-qTOST is uniformly more powerful than the qTOST. Moreover, we also report histograms showing the probability of rejecting H$_0$ along the boundaries of the hypothesis space for the two methods (bottom right panel), which also includes the $\btheta$ parameter at which we evaluate the test size. This result confirms the conservativeness of qTOST, whose empirical size remains close to 3\% and below the simulation error tolerance (displayed as a grey region around $\alpha = 5\%$), while the $\alpha$-qTOST accurately controls the type I error and its empirical test size never exceeds such a tolerance. 

\section{Case studies} 
\label{sec:case_study}

In this section, we apply the qTOST and $\alpha$-qTOST on two case studies. 
In Section~\ref{sec:case_study_1}, we consider a bridging clinical trial between male and
female populations for an HIV treatment when the focus is on a single quantile.
In Section~\ref{sec:case_study_2}, we perform an analysis to compare the reproducibility of an identical experimental protocol performed by different operators in the context of topical products when jointly comparing two quantiles.

\subsection{Case study 1: A bridging clinical trial across gender populations} 
\label{sec:case_study_1}

In this section, we analyze a bridging clinical trial from male to female HIV-positive patients. 
The study, with $n_x=106$ male and $n_y=14$ female patients, examined the co-administration of tipranavir/ritonavir (TPV/r), given twice-daily with an oral dosage of 500 and 200 mg, respectively.
The data were collected from a United States Food and Drug Administration drug label, and are publicly available at \url{https://www.accessdata.fda.gov/drugsatfda_docs/label/2024/021814s030lbl.pdf}
(see Table 5 therein).
The primary objective of this bridging study, which was inspired by the case study presented in \citet{pei2008statistical}, was to evaluate whether the TPV pharmacokinetics upon co-administration with ritonavir resulted in comparable blood concentrations between female patients and male patients.
The PK parameter of interest is plasma trough concentration (Cp$_{\text{trough}}$), which plays a critical role in the efficacy of drugs that require sustained minimum plasma levels, such as antibiotics, antivirals, and immunosuppressants. 
In the case of TPV/r, maintaining adequate trough concentrations of TPV is particularly important for ensuring antiviral efficacy.
Therefore, we focus on assessing equivalence on the lower tail of the (log-transformed) Cp$_{\text{trough}}$ distribution, such as the 15\% or 20\% percentiles, since patients with lower drug exposure may experience treatment failure. 
\begin{table}[t]
\hspace*{-1.5cm}
\centering
\footnotesize
\renewcommand{\arraystretch}{1.4}
\begin{tabular}{llccldddd}
\hline
\multirow{2}{*}{Data} & \multirow{2}{*}{Compound} & \multirow{2}{*}{$n_x$} & \multirow{2}{*}{$n_y$} & \multirow{2}{*}{Parameter} & \multicolumn{2}{c}{$X$ (reference)} & \multicolumn{2}{c}{$Y$ (target)} \\
\cline{6-9}
& & & &  & \multicolumn{1}{c}{Mean} & \multicolumn{1}{c}{SD} & \multicolumn{1}{c}{Mean} & \multicolumn{1}{c}{SD}  \\
\hline
HIV  & Tipranavir & 106 & 14 & Cp$_{\text{trough}}$ ($\mu$M) & 35.6 & 16.7 & 41.6 & 24.3 \\
& & &  & $\log($Cp$_{\text{trough}}$$)$ & 3.47 & 0.45 & 3.58 & 0.54 \\
Cutaneous biodistribution  & Molecule X & 6 & 6 & Amount (ng/cm$^2$) & 251.39 & 149.32 & 226.93 & 83.34\\
& & &  & $\log$(Amount)  & 5.40 & 0.54 & 5.36 & 0.40\\
\hline
\end{tabular}
\caption{Sample size, mean and standard deviation (SD) on the original and log-transformed scales for the data used in the two case studies.}
\label{table:cs_desc_stats}
\end{table}

Table \ref{table:cs_desc_stats} presents descriptive statistics for the Cp$_{\text{trough}}$ PK parameter of interest across the two groups of patients.
The PK parameters demonstrate substantial variability. Assuming a log-normal distribution for the Cp$_{\text{trough}}$, as customary in these applications \citep{pei2008statistical}, we applied a moment-based transformation to better approximate normality. 
Namely, 
using the subscript $o$ to denote estimates obtained on data in the original scale, 
we consider
$
\barx = \log \{ \overline{X}_o^2  (\overline{X}_o^2+\widehat{\sigma}_{x_o}^2)^{-1/2} \}
$
and
$
\hsigmax^2 = \log (1+ \widehat{\sigma}_{x_o}^2 / \overline{X}_o^2 ) , 
$
and similarly construct $\bary$ and $\hsigmay$.
For our analyses, to assess the hypothesis in \eqref{eq:hyp_pi}, we employ these transformed summary statistics, which are reported on the second row of Table~\ref{table:cs_desc_stats}.

\begin{figure}[t]
    \centering
    \includegraphics[width=1\textwidth]{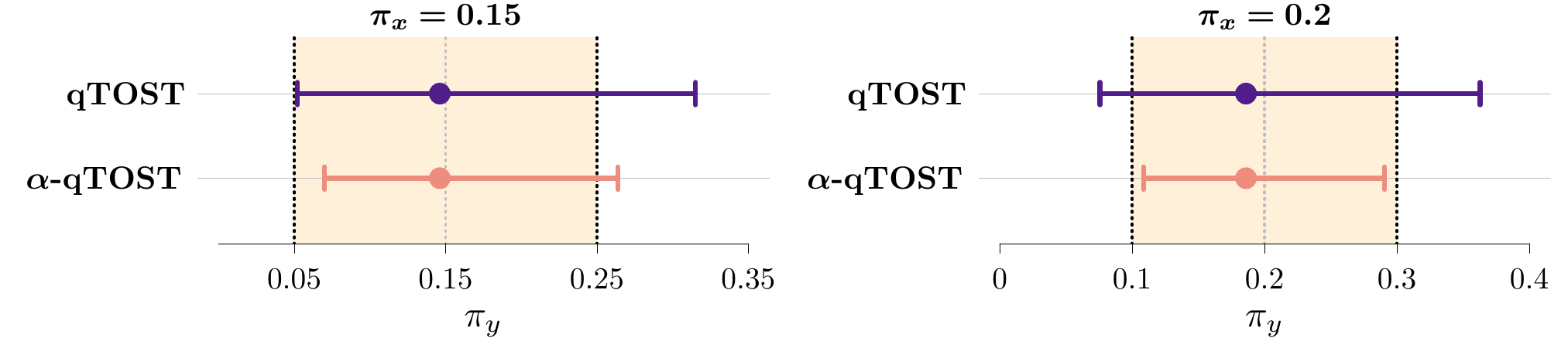}
\caption{Confidence intervals for $\pi_y$ at the level 100($1-2\alpha$)\%  for the qTOST and $\alpha$-qTOST procedures at the two quantiles of interest in the HIV dataset: $\pix = 0.15$ (left panel) and $\pix = 0.20$ (right panel). 
Dashed vertical black lines correspond to equivalence margins $\pi_x \pm c$ with $c = 0.1$. Equivalence can be declared for a method if its confidence interval is entirely contained within the orange region $(\pi_x-c, \pi_x+c)$.}
\label{fig:equiv_CIs}
\end{figure}
Figure \ref{fig:equiv_CIs} presents the 100($1-2\alpha$)\% confidence intervals obtained by the qTOST and $\alpha$-qTOST procedures for the two scenarios of interest: $\pix = 0.15$ (left panel) and $\pix = 0.2$ (right panel). 
Here we fix $\alpha = 5\%$ and $c=10\%$, considering the latter as an appropriate threshold for establishing therapeutic equivalence while maintaining clinical relevance.
The point estimates are $\htheta \approx -1.053$ and $\hsigma \approx 0.348$. 
Though both approaches fail to declare equivalence in the scenario when $\pix = 0.15$, the $\alpha$-qTOST approach yields comparatively narrower confidence intervals, indicating a larger power for this test procedure. In contrast, the qTOST method produces wider intervals, reflecting its conservativeness. In the setting with $\pix = 0.20$, where $\htheta \approx -0.892$ and $\hsigma \approx 0.329$, while the qTOST confidence interval is [0.076, 0.362] and does not allow us to declare equivalence, the $\alpha$-qTOST adjusts the test level to $\halpha^* \approx 15.03\%$ leading to a confidence interval of [0.109, 0.291], which is entirely contained within the equivalence margins and thus leads to a declaration of equivalence. 

\begin{figure}[t]
    \centering
    \includegraphics[width=1\textwidth]{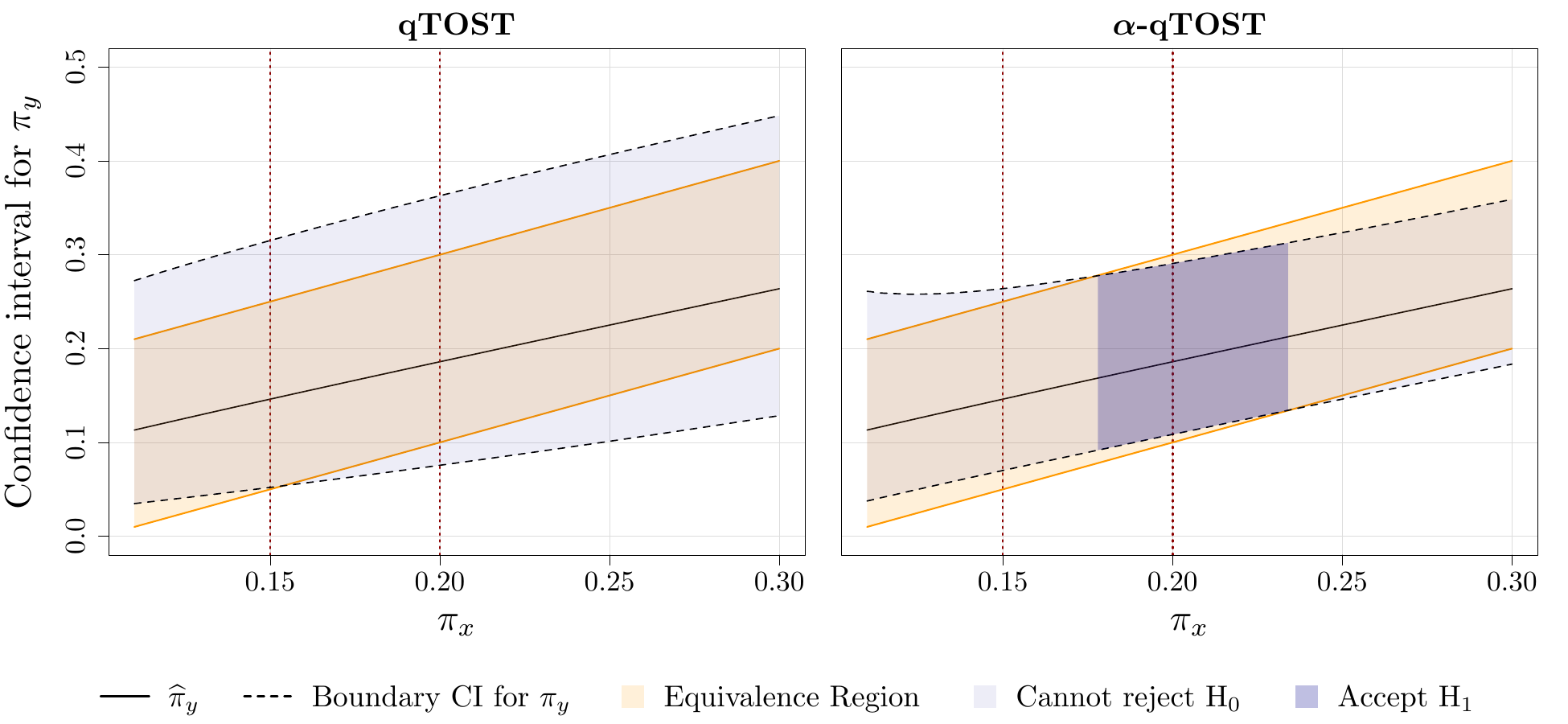}
\caption{Point-wise confidence intervals at the level 100($1-2\alpha$)\% for $\pi_y$ (y-axis) as a function of $\pi_x$ (x-axis) obtained using the qTOST (left panel) and $\alpha$-qTOST (right panel) procedures on the HIV dataset. At any given $\pi_x$, such as $\pi_x \in \{0.15, 0.2\}$ highlighted in dark red, quantile equivalence is established when the confidence interval in lighter blue falls completely within the equivalence region $\pi_x \pm c$ highlighted in orange, with $c = 0.1$. The regions where quantile equivalence can be established are highlighted in darker blue.}
\label{fig:equiv_region}
\end{figure}
While our primary focus is on testing $\pix=0.15$ and $\pix=0.2$, as a further illustration, we independently assess the same hypothesis across a sequence of (lower) $\pix$ quantiles for the transformed data.
Specifically, we construct point-wise confidence intervals using equally-spaced values of $\pi_x \in [0.1, 0.3]$, maintaining $c = 10\%$ and $\alpha=5$\%.
For each parameter $\pi_x$, Figure \ref{fig:equiv_region} illustrates the equivalence regions, constructed as $\pi_x \pm c$ and highlighted in orange, and the point-wise 100($1-2\alpha$)\% confidence intervals for the qTOST (left panel) and $\alpha$-qTOST (right panel) highlighted in blue.
As expected, the $\alpha$-qTOST procedure consistently leads to narrower confidence intervals compared to qTOST.
While qTOST does not lead to a declaration of equivalence at any of the considered $\pi_x$'s, the $\alpha$-qTOST can establish BE for a wide range of $\pi_x$'s (approximately for the percentiles in the range 18-23\%). 
The expanded range of equivalence declaration for $\alpha$-qTOST could further support the efficacy of the drug on the female population while mitigating the risk of developing viral resistance.

\subsection{Case study 2: Demonstrating inter-operator reproducibility}
\label{sec:case_study_2}

In this section, we evaluate the inter-operator reproducibility of an identical experimental protocol for generating biodistribution profiles.
Briefly, the in vitro skin delivery study was performed 
using freshly dermatomed human abdominal skin (with a sample size of 12), under finite dose conditions \citep{oecd2022dermal} using standard Franz diffusion cells.  
A 10 mL solution of the formulation containing the permeant of interest (``molecule X'') was applied to the skin surface (5 mg/cm$^2$).
Upon completion of the experiment (after 16 h), the skin sample was thoroughly cleaned before a central disc (of diameter 8 mm) was punched out, embedded in optimal cutting temperature medium, and snap-frozen in isopentane chilled with liquid nitrogen. 
Then, skin samples from different donors were assigned to two operators, each performing the same experiment with the same number of replicates (i.e.,~$n_x = n_y = 6$), thereby enabling an inter-operator comparison with respect to quantile equivalence. 
However, the results of Operator X were considered as the reference given the greater experience with the experimental technique.  
The skin discs (of diameter 8 mm) were cryosectioned into twenty lamellae, each with a thickness of 20 $\mu$m. The individual lamellae were placed in an Eppendorf tube, and molecule X was extracted using a validated protocol. The extracts were centrifuged and filtered prior to quantification by a validated UHPLC-MS/MS method to determine the amount in each of the lamellae and thereby obtain the spatial distribution profile. 
In this illustration, we restrict our attention to the first 15 lamellae for molecule X, corresponding to skin depths ranging from 0-300 $\mu$m, which encompass anatomic relevant regions including the \textit{stratum corneum}, \textit{viable epidermis}, and \textit{upper dermis}. 
As it is customary to assume log-normality of the original concentration measurements (see e.g.,~\citealt{keene1995log,julious2000pharmacokinetic}), we perform our analysis on the log scale.
Table~\ref{table:cs_desc_stats} presents the summary statistics on the original and the log-transformed scale, and the raw data used for this study are available on the \texttt{cTOST} package in \texttt{R}.

\begin{figure}[t]
    \centering
    \includegraphics[width=1\textwidth]{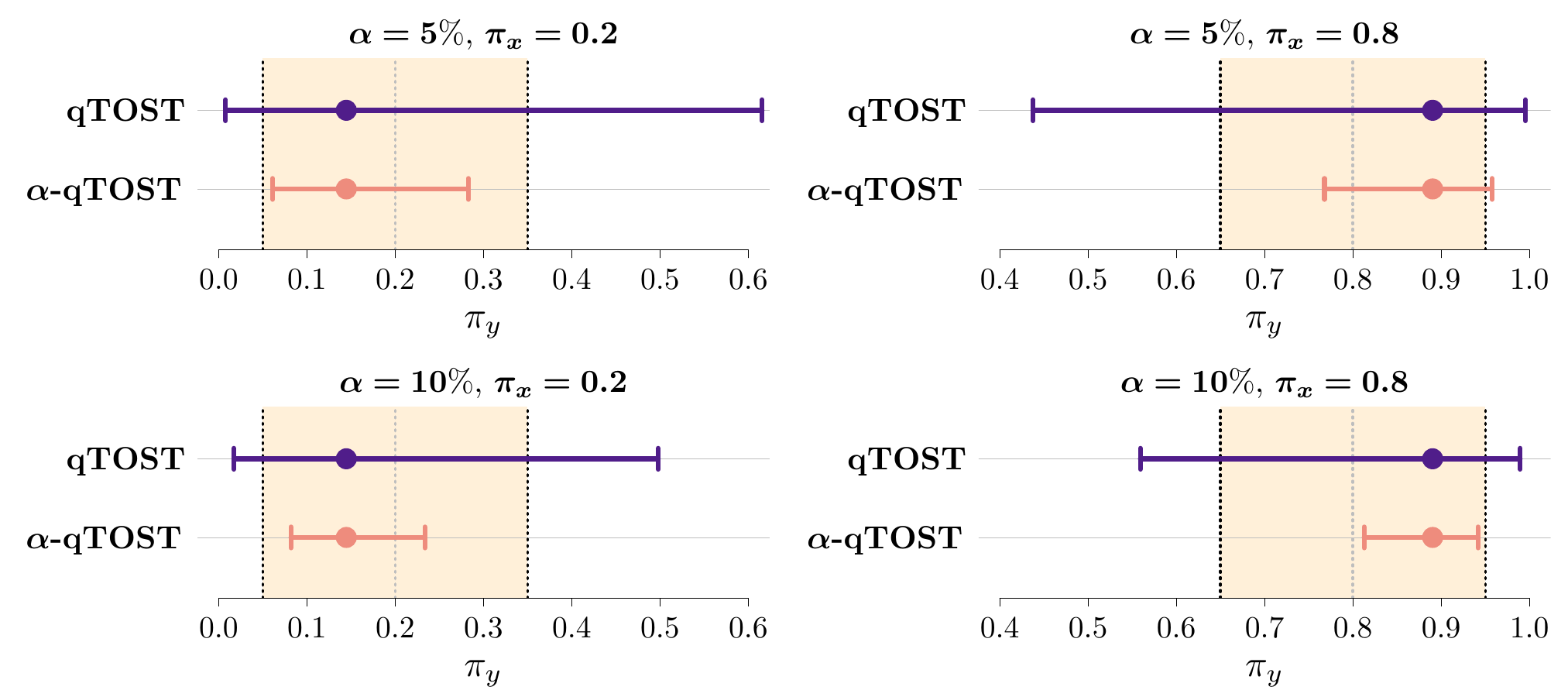}
\caption{
Marginal confidence intervals for $\bpi_y$ at the level 100($1-2\alpha$)\%  for the qTOST and $\alpha$-qTOST procedures when $\pi_{x_1}=0.2$ (first column) and $\pi_{x_2}=0.8$ (second column) for two nominal significance levels of interest in the cutaneous biodistribution dataset: $\alpha = 0.05$ (first row) and $\alpha = 0.10$ (second row). 
Dashed vertical black lines correspond to equivalence margins $\bm{\pi}_x \pm c$ with $c = 0.15$. Equivalence can be declared for a method if both of its confidence intervals are entirely contained within the corresponding orange regions $(\pi_{x_k}-c, \pi_{x_k}+c)$, for $k=1,2$.
}
\label{fig:equiv_CIs_mvt}
\end{figure}

\begin{figure}[t!]
    \centering
    \includegraphics[width=1\textwidth]{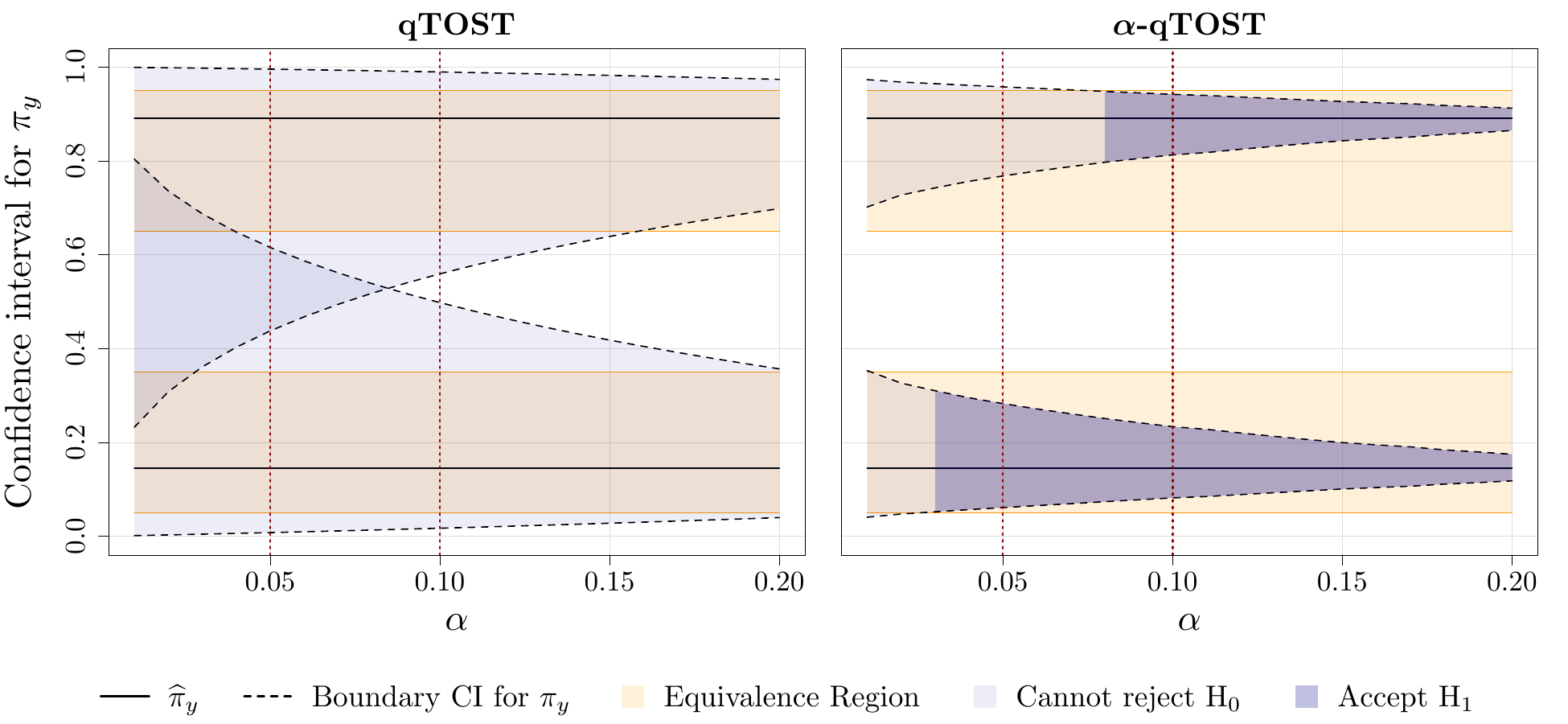}
\caption{
Point-wise confidence intervals at the level 100($1-2\alpha$)\% for $\pi_{y_1}$ and $\pi_{y_2}$ (y-axis) as a function of $\alpha$ (x-axis) obtained using the qTOST (left panel) and $\alpha$-qTOST (right panel) procedures on the cutaneous biodistribution dataset. At any given $\alpha$, such as $\alpha \in \{0.05, 0.1\}$ highlighted in dark red, quantile equivalence is established when both confidence intervals in lighter blue fall completely within the equivalence region $\bm{\pi}_{x} \pm c$ highlighted in orange, with $c = 0.15$. The regions where marginal quantile equivalence can be established are highlighted in darker blue.
}
\label{fig:equiv_region_mvt}
\end{figure}

Thus, we evaluate joint quantile equivalence for the 20 and 80\% percentiles across the 
results generated by the two operators following an identical experimental protocol, under a symmetric equivalence margin of $c=15\%$. 
This enables us to compare the biodistribution profiles generated by each operator. Establishing the equivalence of these profiles demonstrates the reproducibility of the cutaneous biodistribution method independently of the operator conducting the experiment.
The multiple quantiles approach may be of interest when it comes to an evaluation of a product’s safety and efficacy margin and eventual inter-individual variations.
Figure~\ref{fig:equiv_CIs_mvt} displays the 100($1-2\alpha$)\% confidence intervals obtained from the qTOST and $\alpha$-qTOST procedures for the joint assessment of $\pi_{x_1} = 0.2$ (left column) and $\pi_{x_2} = 0.8$ (right column). Specifically, we test both approaches in the scenario when $\alpha = 5\%$ (first row) and $\alpha = 10\%$ (second row).
When $\alpha = 5\%$, both procedures fail to establish joint quantile equivalence.
In particular, while both marginal confidence intervals for qTOST exceed the equivalence margins, only the one associated with $\pi_{x_2}$ does not allow for the declaration of equivalence for the $\alpha$-qTOST.  
However, under a less restrictive nominal level $\alpha=10$\%, the $\alpha$-qTOST procedure adjusts the significance level at $\halpha^* = 34.15\%$, which results in narrower confidence intervals. In this case, the adjusted procedure leads to a declaration of BE, whereas the qTOST does not.

As an illustration, we jointly assess equivalence at these quantiles using both qTOST and $\alpha$-qTOST across a sequence of equally-spaced values of $\alpha \in [0.01, 0.2]$. Figure~\ref{fig:equiv_region_mvt} reports the corresponding $100(1-2\alpha)\%$ point-wise confidence intervals, where we maintain $c = 15\%$. This shows that, at any given nominal significance level $\alpha$, the qTOST confidence interval for any $\pi_{y_k}$ extends beyond the equivalence margins, resulting in non-rejection of H$_0$. In contrast, the $\alpha$-qTOST procedure provides a less conservative alternative to qTOST, yielding narrower confidence intervals and leading to a declaration of quantile equivalence for a wide range of nominal $\alpha$ levels (approximately for levels in the range 8-20\%).

\section{Final remarks}
\label{sec:final}

In this article, we extended the quantile BE framework of \cite{pei2008statistical} for testing a single quantile between two normal populations
to the simultaneous assessment of multiple quantiles. To address the conservativeness of the traditional qTOST procedure, we introduced the $\alpha$-qTOST adjustment. By adjusting the test size to match the nominal significance level, the proposed method achieves uniformly larger power compared to qTOST while maintaining control of the type I error. The advantages of $\alpha$-qTOST are more pronounced in scientifically relevant scenarios, such as heteroskedastic settings with unbalanced sample sizes and when testing more extreme quantiles. In addition, a computationally efficient algorithm to construct the $\alpha$-qTOST adjustment was proposed, making it practical for routine use.

The proposed methodology addresses critical challenges in clinical and pre-clinical research, particularly in the design and analysis of bridging studies.
Such studies aim to extrapolate the findings from one (well-studied) reference group to support drug or experimental protocol approval in a target (often under-represented) group. Here it is often impractical or unethical to conduct a full clinical trial on the target population, leading to much smaller sample sizes and/or more noisy data. This is particularly important when examining therapeutic effects across different ethnic groups, age ranges, or other demographic factors where drug response patterns may vary systematically.
Comparing an extreme quantile is valuable in the context of systemic drugs, as this can be linked to safety and efficacy concerns. More specifically, lower quantiles are critical for ensuring therapeutic efficacy, as inadequate drug exposure may lead to treatment failure or the development of resistance, while upper quantiles are essential for evaluating safety margins and potential toxicity risks.
We considered one such case study related to HIV drug development between male and female patient populations, where the evaluation at low PK levels is important in minimizing the risk of the development of drug resistance \citep{little2002antiretroviral,soeria2024sub}.
Moreover, the joint assessment of multiple quantiles enables a more comprehensive evaluation of drug response distributions, for instance, when comparing the therapeutic index of a drug through the joint examination of efficacy-related lower quantiles and safety-related upper quantiles. This provides deeper insights into population-specific profiles and supports more informed regulatory decision-making.
Finally, we considered a bridging case study to assess equivalence in experimental protocols performed by different operators in the context of locally acting drugs. 
To date, the ``cutaneous biodistribution method'' has principally been used in the development and optimization of pharmaceutical formulations of poorly water-soluble drugs with dermatologic indications \citep{lapteva2014polymeric,lapteva2019self,kandekar2019polymeric,quartier2021polymeric,darade2023polymeric}, and proposed as an innovative and data-rich approach to establish BE of locally acting, topically applied formulations \citep{quartier2019cutaneous}. 
The joint assessment of multiple quantiles allowed us to validate the reproducibility of the cutaneous biodistribution method following an identical experimental protocol performed by different operators. This is another step in showing how the biodistribution profiles could serve as an innovative tool for establishing BE of topically applied locally acting formulations, potentially contributing to the approval of generic drugs.

\bibliography{refs.bib}

\cleardoublepage
\newpage

\bigskip
\begin{center}
{\Large\bf APPENDIX}
\end{center}

\appendix
\renewcommand{\thetable}{A.\arabic{table}}
\setcounter{table}{0}
\renewcommand{\theequation}{A.\arabic{equation}}
\setcounter{equation}{0}
\renewcommand{\thefigure}{A.\arabic{figure}}
\setcounter{figure}{0}

\section{Theoretical results}

\subsection{Goodness of the approximation}
\label{app:omega_tilde_closeness}

In this section, we demonstrate the convergence rate for $\htheta$ in \eqref{eq:mle_theta}. 
Namely, we show that 
\be \label{eq:theta_tilde_rate}
    \htheta = \wt\theta + O_p(n^{-1}) .
\ee
Therefore, since $\wt\theta = \theta + O_p(n^{-1/2})$ (see e.g.,~\citealt{boulaguiem2024finite}), we have that in large samples $\htheta$ and $\wt\theta$ are much closer than $\htheta$ and $\theta$.
In turn, this suggests that the properties derived for $\wt\alpha^*$ in \eqref{eq:alpha_star_tilde} extend to $\alpha^*$ in \eqref{eq:alpha_star} as the sample size increases.

Let 
$$
W_y \equiv \sqrt{ \nu_y \frac{\hsigmay^2}{\sigma_y^2} } ,
$$
so that 
\begin{align*}
    \mathbb{E} (W_y) = \sqrt{\nu_y} + O(n_y^{-1}) \quad \text{and} \quad \var(W_y) = \frac{\nu_y}{2 n_y} \{ 1 + O(n_y^{-1}) \} = \frac{1}{2} + O(n_y^{-1}) .
\end{align*}
Moreover, let $a_y^2 = \lim_{n_y \to \infty} n_y \sigma_y^2$, and 
$a_x^2 = \lim_{n_x \to \infty} n_x \sigma_x^2$,
Then, based on \eqref{eq:mle_sigmas}, we get
\begin{align}
    \hsigmay &= W_y \frac{\sigmay}{\sqrt{\nu_y}} = 
    \frac{\sigmay}{\sqrt{\nu_y}} \left[ \left\{ \frac{W_y - \mathbb{E}(W_y)}{\sqrt{\var(W_y)}} \right\} \sqrt{\var(W_y)} + \mathbb{E}(W_y) \right] \\
    &=
    \sigma_y + \frac{a_y}{n_y} \left\{ O_p(1) + O(n_y^{-1}) \right\} = 
    \sigma_y + O(n_y^{-1}) ,
\end{align}
and similarly
$$
    \hsigmax = \sigma_x + O(n_x^{-1}) .
$$
Due to $n_y (\hsigmay - \sigma_y) = O_p(1)$, we also have 
$$
    \hsigmay - \sigma_y = \frac{a_y}{2 n_y } \frac{W_y - \mathbb{E}(W_y)}{\sqrt{\var(W_y)}} + O(n_y^{-2})  \asymp_p \frac{1}{n_y} .
$$
Then, based on a Taylor expansion, we get 
\be \label{eq:theta_tilde_order1}
    \frac{\hsigmax}{\hsigmay} = \frac{\sigmax}{\sigmay} + O_p ( n^{-1} ) ,
\ee 
and similarly
\be \label{eq:theta_tilde_order2}
    \frac{\bar{X} - \bar{Y}}{\hsigma_y} = \frac{\bar{X} - \bar{Y}}{\sigma_y} + O_p\left( \frac{\bar{X} - \bar{Y}}{n_y} \right) = 
    \frac{\bar{X} - \bar{Y}}{\sigma_y} + O_p\left( n^{-1/2} n_y^{-1} \right) .
\ee 
Combining \eqref{eq:theta_tilde_order1} and \eqref{eq:theta_tilde_order2}, we obtain
$$
    \htheta = \frac{\bar{X} - \bar{Y}}{\hsigma_y} + \frac{\hsigma_x}{\hsigma_y} \Phi^{-1}(\pi_x) = 
    \frac{\bar{X} - \bar{Y}}{\sigma_y} + O_p\left( n^{-1/2} n_y^{-1} \right) + \frac{\sigmax}{\sigmay} + O_p ( n^{-1} ) =
    \wt\theta + O_p ( n^{-1} ) , 
$$
which verifies \eqref{eq:theta_tilde_rate}.

\subsection{Conservativeness of qTOST}
\label{app:TOST_conservativeness}

In this section, we illustrate the conservative nature of the qTOST procedure based on \eqref{eq:tost_power_known_sigmas}. 
Without any loss of generality, we consider equivalence margins in \eqref{eq:hyp_theta} that are symmetric around zero, by taking $\delta \equiv \delta_u = - \delta_l$.
Assume that $\tau^2 < D^2_{\delta, \alpha}$ where $D_{\delta, \alpha}$ is chosen such that 
\be \label{eq:app_D_bound}
- \delta + z_\alpha \tau < \delta - z_\alpha \tau \Longleftrightarrow \tau < \frac{\delta}{z_\alpha} \equiv D_{\delta, \alpha} .
\ee
Thus, based on \eqref{eq:tost_power_known_sigmas}, we have  
\bse
\begin{aligned}
    \wt\omega(\theta, \tau, \alpha) 
    &= \Pr\left (-\delta + z_\alpha \tau <\wt\theta< \delta -z_\alpha \tau \; \big\lvert \; \theta, \tau, \alpha \right) \\
    &= \Pr\left (z_\alpha - \frac{\delta + \theta}{\tau} < Z < -z_\alpha + \frac{\delta - \theta}{\tau} \; \Big\lvert \; \theta, \tau, \alpha \right) ,
\end{aligned}
\ese 
where $Z \sim \mathcal{N}(0, 1)$.
Moreover, since $\tau < D_ {\delta, \alpha}$, it follows that 
$
z_\alpha - \frac{\delta + \theta}{\tau} < -z_\alpha + \frac{\delta - \theta}{\tau}
$
for any $\theta \in \real$.
Letting $v \equiv -z_\alpha + \frac{\delta}{\tau} > 0$, where the inequality is due to \eqref{eq:app_D_bound}, we obtain 
\be \label{eq:app_power_even}
    \wt\omega(\theta, \tau, \alpha) 
    = \Phi\left(v - \frac{\theta}{\tau}\right) - \Phi\left(-v - \frac{\theta}{\tau}\right) 
    = \Phi\left(v - \frac{\theta}{\tau}\right) + \Phi\left( v + \frac{\theta}{\tau}\right) - 1 ,
\ee
demonstrating that $\wt\omega(\cdot)$ is an even function of $\theta$, in the sense that $\wt\omega(\theta, \tau, \alpha) = \wt\omega(-\theta, \tau, \alpha)$.
To study the size of the test, we start by considering the partial derivative of $\wt\omega(\theta, \tau, \alpha)$ with respect to $\theta$: 
\bse 
    \frac{\partial}{\partial \theta} \wt\omega(\theta, \tau, \alpha) = \frac{1}{\tau} \left\{ \varphi\left( v + \frac{\theta}{\tau} \right) - \varphi\left( v - \frac{\theta}{\tau} \right) \right\} ,
\ese
where $\varphi(\cdot)$ denotes the probability density functions of a standard normal random variable.
Due to \eqref{eq:app_power_even}, we can simply study the case for $\theta>0$, where it follows that
\be \label{eq:app_power_decreasing} 
    \left\lvert v + \frac{\theta}{\tau} \right\rvert > \left\lvert v - \frac{\theta}{\tau} \right\rvert
    \Longrightarrow \varphi\left( v - \frac{\theta}{\tau} \right) > \varphi\left( v + \frac{\theta}{\tau} \right) 
    \Longrightarrow \frac{\partial}{\partial \theta} \wt\omega(\theta, \tau, \alpha) < 0 .
\ee
Therefore, returning to the size, combining \eqref{eq:app_power_even} and \eqref{eq:app_power_decreasing} we obtain 
\bse 
\begin{aligned}
    \sup_{\theta \not\in \Theta_1} \, \wt\omega(\theta, \sigma, \alpha) &= \sup_{\theta \geq \delta} \, \wt\omega(\theta, \sigma, \alpha) = \wt\omega(\delta, \sigma, \alpha) \\
    &= \Phi(-z_\alpha) - \Phi\left(z_\alpha - \frac{2\delta}{\tau} \right) \\
    &= \alpha - \Phi\left\{ z_\alpha - \frac{2\delta}{\tau} \right\} = \zeta < \alpha ,
\end{aligned}
\ese
implying that the qTOST procedure based on \eqref{eq:tost_power_known_sigmas} is only level $\alpha$.
However, assuming that $\lim_{n_x,n_y \to \infty } \; \frac{\max(n_x,n_y)}{n_x n_y} = 0$, we have that $\lim_{n_x,n_y \to \infty}\; \zeta = \alpha$ since $\lim_{n_x,n_y \to \infty} \Phi \left\{ z_{\alpha} - \frac{2\delta}{\tau} \right\} = 0$. Thus, the test is asymptotically size-$\alpha$. 

\subsection{Conditions for the existence of $\wt\alpha^*$}
\label{app:alpha_star_existance}

In this section, we provide the conditions for the existence and uniqueness of the size-$\alpha$ adjustment $\wt\alpha^*$ in \eqref{eq:alpha_star_tilde}. To simplify the notation, let $ \wt\omega(\alpha) \equiv \sup_{\theta \not\in \Theta_1} \wt\omega(\theta, \tau, \alpha )$. Consider the set of potential adjustments $\mathcal{A} \equiv \{ x \in [\alpha, 0.5) \, \big|  \, \wt\omega(x)>0 \}$. 
Since
\begin{equation} \label{eq:app_deriv_omega}
\begin{aligned}
\wt\omega' (\xi)  \equiv \frac{\partial}{\partial x} \wt\omega(x) \Big \rvert_{x = \xi} &=  \frac{d}{d \xi} \left[ \xi -  \Phi \left\{
\frac{\Phi^{-1}\left(\pi_{x} - c\right) + z_{\xi}\tau -\Phi^{-1}(\pi_{x} + c) }{\tau}  \right \} \right] \\
&= 1 + \phi \left ( z_\xi - \frac{ 2 \delta }{\tau} \right) \frac{1}{\phi(z_\xi)} > 1 ,
\end{aligned}
\end{equation}
it follows that $\wt\omega(\xi)$ is continuously differentiable and strictly increasing in $\xi$ for $\xi \in \mathcal{A}$.
Then, as $\alpha \geq \wt\omega(\alpha)$ due to \eqref{eq:tost_size}, we have
\begin{align*}
\alpha < \alpha_{\max} &\equiv \lim_{\alpha \to 0.5^-} \wt\omega(\alpha) \\
&= \lim_{\alpha \to 0.5^-} \left[ \alpha - \Phi \left\{ \frac{\Phi^{-1}\left(\pi_{x} - c\right) + z_{\alpha}\tau - \Phi^{-1}(\pi_{x} + c) }{\tau}  \right\} \right] \\
&= \frac{1}{2} - \Phi \left \{ \frac{\Phi^{-1}\left(\pi_{x} - c\right) - \Phi^{-1}(\pi_{x} + c) }{\tau}  \right\}  \\
&= \Phi \left\{ \frac{ \Phi^{-1}(\pi_{x} + c) - \Phi^{-1}\left(\pi_{x} - c\right) }{\tau}  \right\}  - \frac{1}{2} .
\end{align*}
Therefore, for
$$
\tau < \frac{ \Phi^{-1}(\pi_{x} + c) - \Phi^{-1}\left(\pi_{x} - c\right) }{\Phi^{-1}(\alpha + 0.5)} 
$$
we have that $\alpha < \alpha_{\max}$, which ensures that $\wt\alpha^*$ exists and is unique.

\subsection{Convergence rate of the iterative algorithm for $\wt\alpha^*$}
\label{app:algo}

This section demonstrates that the proposed iterative algorithm converges exponentially fast to the target $\wt\alpha^*$ when solving the matching in \eqref{eq:alpha_star_tilde}.
Based on \eqref{eq:app_deriv_omega}, for any $\xi \in \mathcal{A}$, note that $\wt\omega(\xi)$ is continuously differentiable and satisfies 
$$
1 < \wt\omega' (\xi) < 2 ,
$$
where the second inequality is due to $\phi(z_\xi) > \phi(z_\xi - 2 \delta / \tau)$.
The rest of this proof follows the argument in \citet{boulaguiem2024finite}, which we present below for completeness.
Let 
$$
T(\xi) \equiv \alpha + \xi - \wt\omega(\xi) .
$$
Then, for any $\alpha_1, \alpha_2 \in \mathcal{A}$, it follows from the mean-value theorem that 
$$
T (\alpha_1) - T (\alpha_1) = \alpha_1 - \alpha_2 -\wt\omega(\alpha_1) + \wt\omega(\alpha_2) = \alpha_1 - \alpha_2 -\wt\omega'(\alpha_3) (\alpha_2 - \alpha_1) ,
$$
where $\alpha_3 \equiv \xi \alpha_1 + (1-\xi) \alpha_2$ with $\xi \in [0, 1]$. 
Hence
$$
\Big \lvert T (\alpha_1) - T (\alpha_2) \Big \lvert = \Big \lvert (\alpha_1 - \alpha_2)(1-\wt\omega'(\alpha_3)) \Big \rvert < \Big \rvert \alpha_1 - \alpha_2 \Big \rvert .
$$
Based on Kirszbraun theorem \citep{federer2014geometric}, the function $T(\xi)$ can be extended with respect to $\xi \in \mathcal{A}$ to a contraction map from $\real$ to $\real$,
and Banach fixed-point theorem ensures that the sequence $T(\wt\alpha^{*(k)})$ converges as $k \to \infty$.
Let $\wt\alpha^*$ be the limit of the sequence $ \{\wt\alpha^{*(k+1)}\}_{k \in \integer}$ which, by construction, is the unique fixed point of the function $T(\xi)$.
Therefore, 
$$
\wt\alpha^* = T(\wt\alpha^*) = \alpha + \wt\alpha^* - \wt\omega(\wt\alpha^*) .
$$
Rearranging terms provides
$$
\wt\alpha^* = \argzero_{\xi \in \mathcal{A}}  \left\{ \wt\omega(\xi) - \alpha \right\} = \argzero_{\xi \in [\alpha,0.5)}  \left\{ \wt\omega(\xi) - \alpha \right\} ,
$$
which ensures the convergence of the sequence $ \{\wt\alpha^{*(k+1)}\}_{k \in \integer}$.
This implies the existence of some $0 < \epsilon < 1$ such that for $k \in \integer$ we obtain
$$
\Big\lvert \wt\alpha^{*(k+1)} - \wt\alpha^* \Big\rvert < \epsilon^k \Big\lvert \wt\alpha^* - \alpha \Big\rvert < \frac{1}{2} e^{-bk} ,
$$
for some constant $b>0$.

\subsection{Extension to multiple quantiles}
\label{app:mvt}

In this section, we derive the asymptotic covariance matrix $\bSigma_a$ in \eqref{eq:asymp_theta_dist_mvt}. Without loss of generality, we only consider the case where $K=2$ for two arbitrary quantiles of interest $\pi_{x_1}$ and $\pi_{x_2}$.
Let 
$\bdeta \equiv [\mu_x-\mu_y , \sigma_x, \sigma_y]^T$ and 
$\hbdeta \equiv [\barx-\bary , \hsigma_x,  \hsigma_y]^T$, and define
$$
\theta_i = f(\bdeta, D_i) \equiv \frac{\mu_x - \mu_y}{\sigma_y} + \frac{\sigma_x}{\sigma_y} D_i \quad \text{and} \quad 
\htheta_i = f(\hbdeta, D_i) \equiv \frac{\barx - \bary}{\hsigma_y} + \frac{\hsigma_x}{\hsigma_y} D_i ,
$$
where 
$D_i \equiv \Phi^{-1}(\pi_{x_i})$, for $i=1, 2$, is used to simplify the notation.
Assuming that 
$n_y / n_x \asymp 1$ with $ n \asymp n_y \asymp n_x$ such that $n_x / n \to c_1$ and $n_y / n \to c_2$, it follows that 
$$
\sqrt{n} ( \hbdeta - \bdeta) \todist \mathcal{N}(\0, \bOmega) , 
$$
where $\bOmega$ is a diagonal matrix with entries
\begin{align*}
\lim_{n \to \infty} \Omega_{1,1} &= \lim_{n \to \infty} \var\{\sqrt{n} (\barx - \bary)\} = \lim_{n \to \infty} \left( \frac{n}{n_x} \sigma_x^2 + \frac{n}{n_y} \sigma_y^2 \right) = 
\frac{\sigma_x^2}{c_1} + \frac{\sigma_y^2}{c_2} \\
\lim_{n \to \infty} \Omega_{2,2} &= \lim_{n \to \infty} \var(\sqrt{n} \hsigma_x^2) = \lim_{n \to \infty} \frac{n}{\nu_x} \sigma_x^2 \var\left(\sqrt{\nu_x \frac{\hsigma_x^2}{\sigma_x^2} }\right) = \lim_{n \to \infty} \left\{ \frac{n}{2 n_x} \sigma_x^2 + O(n^{-1}) \right\} = \frac{\sigma_x^2 }{2 c_1}  \\
\lim_{n \to \infty} \Omega_{3,3} &= \lim_{n \to \infty} \var(\sqrt{n} \hsigma_y^2)  = \frac{\sigma_y^2 }{2 c_2} .
\end{align*} 

By a Taylor expansion, we get
\begin{align*}
    \sqrt{n} (\htheta_i - \theta_i) = \sqrt{n} \{ f(\hbdeta, D_i) - f(\bdeta, D_i)  \} 
    =  \nabla f(\bdeta, D_i) \sqrt{n} (\hbdeta - \bdeta) + o_p(1) , 
\end{align*}
using the continuous mapping theorem and where 
\begin{align*}
    \nabla f(\bdeta, D_i) \equiv \frac{\partial}{\partial \bdeta^T} f(\bdeta, D_i) 
    &= \left[ 
    \frac{\partial}{\partial (\mu_x- \mu_y)}  f(\bdeta, D_i) ,
    \frac{\partial}{\partial \sigma_x}  f(\bdeta, D_i) ,
    \frac{\partial}{\partial \sigma_y}  f(\bdeta, D_i)
    \right]^T \\
    &= \left[ 
    \frac{1}{\sigma_y} ,
    \frac{D_i}{\sigma_y} ,
    - \left( \frac{\mu_x - \mu_y}{\sigma_y^2} + \frac{\sigma_x}{\sigma_y^2} D_i \right)
    \right]^T .
\end{align*} 
Therefore, we have 
\begin{align*}
    \lim_{n \to \infty} \cov \{ & \sqrt{n} ( \htheta_1 - \theta_1) , \sqrt{n} ( \htheta_2 - \theta_2) \} =  \nabla f(\bdeta, D_i)^T \bOmega \nabla f(\bdeta, D_i) \\
    &=
    \frac{1}{\sigma_y^2} \left[ \frac{\sigma_x^2}{c_1} + \frac{\sigma_y^2}{c_2}  + \frac{D_1 D_2 \sigma_x^2}{2 c_1} + \frac{1}{2 c_2}\{ (\mu_x-\mu_y) + \sigma_x D_1 \} \{ (\mu_x-\mu_y) + \sigma_x D_2 \} \right] \\
    &=
     \frac{1}{2 c_2} \theta_1 \theta_2 +
     \frac{(D_1 D_2 +2)}{2 c_1} \frac{\sigma_x^2}{\sigma_y^2} + \frac{1}{c_2} \\
     &=
     \frac{1}{c_2} \left[ 1 + \frac{\theta_1 \theta_2}{2}  +
     \frac{l}{\gamma} \left\{ 1 + \frac{ D_1 D_2}{2} 
     \right\} \right] ,
\end{align*}
and as an estimator for $\cov(\htheta_1 , \htheta_2)$ we obtain
$$  
     \wh\cov(\htheta_1 , \htheta_2) = \frac{1}{n} \left[ 1 + \frac{\htheta_1 \htheta_2}{2}  +
     \frac{l}{\wh\gamma} \left\{ 1 + \frac{ D_1 D_2}{2} 
     \right\} \right] .
$$
Finally, as a special case, note that by taking $\htheta = \htheta_1 = \htheta_2$ we obtain $\wh\cov(\htheta, \htheta) = \wh\var(\htheta) = \wh\sigma^2$ described in \eqref{eq:theta_hat_asym_dist}, which corresponds to the estimator in Eq.~(4) of \citet{pei2008statistical}.

\renewcommand{\thetable}{B.\arabic{table}}
\setcounter{table}{0}
\renewcommand{\theequation}{B.\arabic{equation}}
\setcounter{equation}{0}
\renewcommand{\thefigure}{B.\arabic{figure}}
\setcounter{figure}{0}

\section{Additional simulation results}

\subsection{Testing a single quantile}
\label{app:simul}

\begin{figure}[ht!]
    \centering
    \includegraphics[width=1\textwidth]{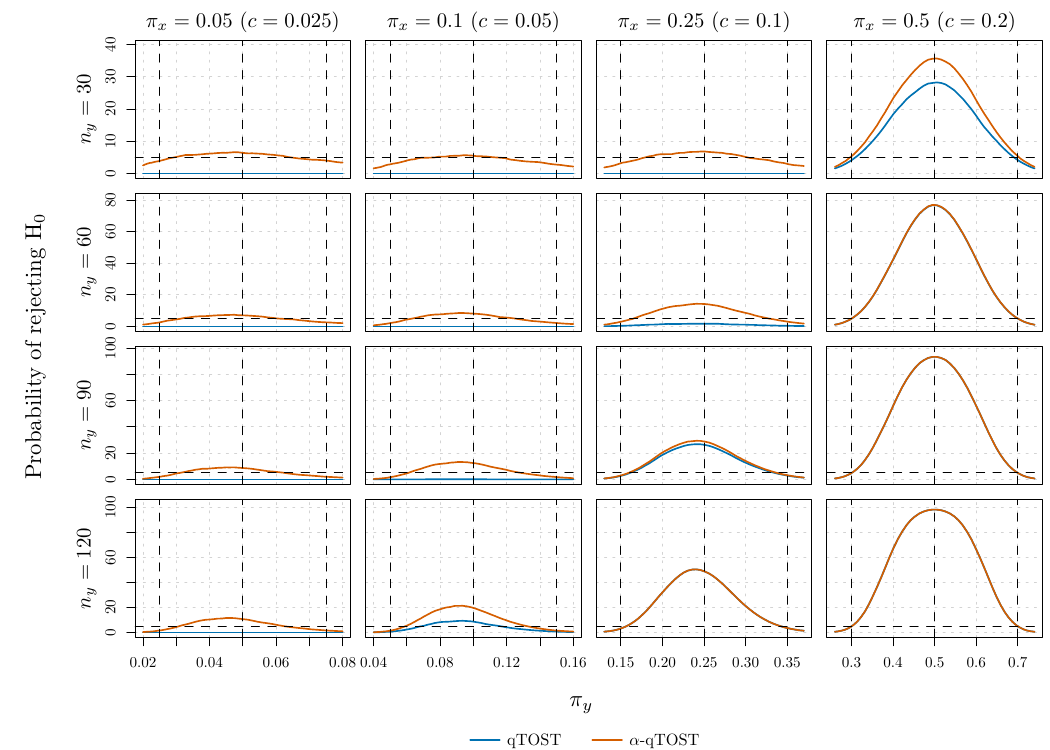}
    \caption{Simulation results comparing the probability of rejecting H$_0$ for $\alpha$-qTOST and qTOST when $\gamma=1$ and $l=1$.}
    \label{fig:power_sim_setting1_gamma_1_l_1}
\end{figure}

\begin{figure}[ht!]
    \centering
    \includegraphics[width=1\textwidth]{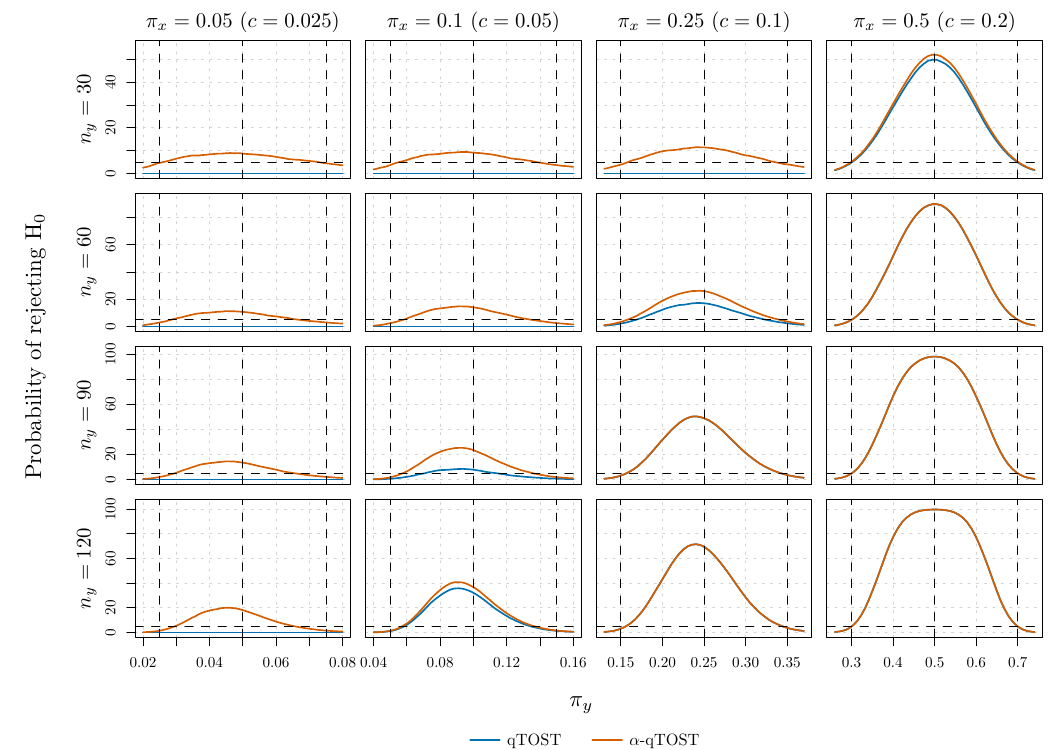}
    \caption{Simulation results comparing the probability of rejecting H$_0$ for $\alpha$-qTOST and qTOST when $\gamma=1$ and $l=\nicefrac{1}{2}$.}
    \label{fig:power_sim_setting1_gamma_1_l_2}
\end{figure}

\begin{figure}[ht!]
    \centering
    \includegraphics[width=1\textwidth]{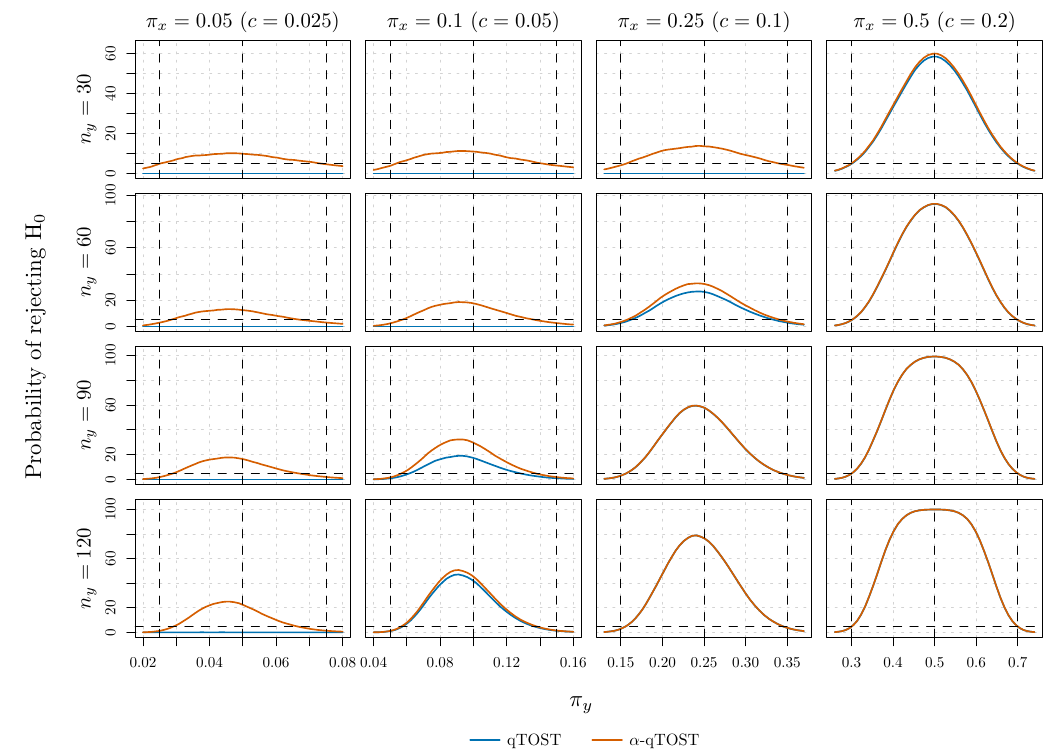}
    \caption{Simulation results comparing the probability of rejecting H$_0$ for $\alpha$-qTOST and qTOST when $\gamma=1$ and $l=\nicefrac{1}{3}$.}
    \label{fig:power_sim_setting1_gamma_1_l_3}
\end{figure}

\begin{figure}[ht!]
    \centering
    \includegraphics[width=1\textwidth]{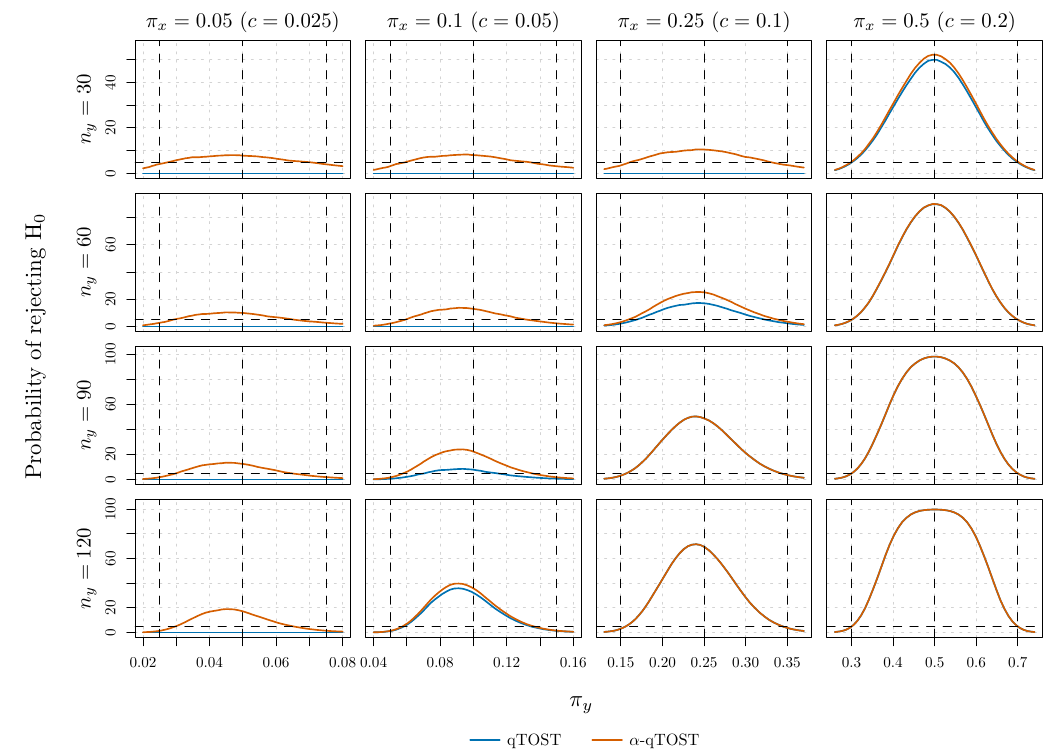}
    \caption{Simulation results comparing the probability of rejecting H$_0$ for $\alpha$-qTOST and qTOST when $\gamma=2$ and $l=1$.}
    \label{fig:power_sim_setting1_gamma_2_l_1}
\end{figure}

\begin{figure}[ht!]
    \centering
    \includegraphics[width=1\textwidth]{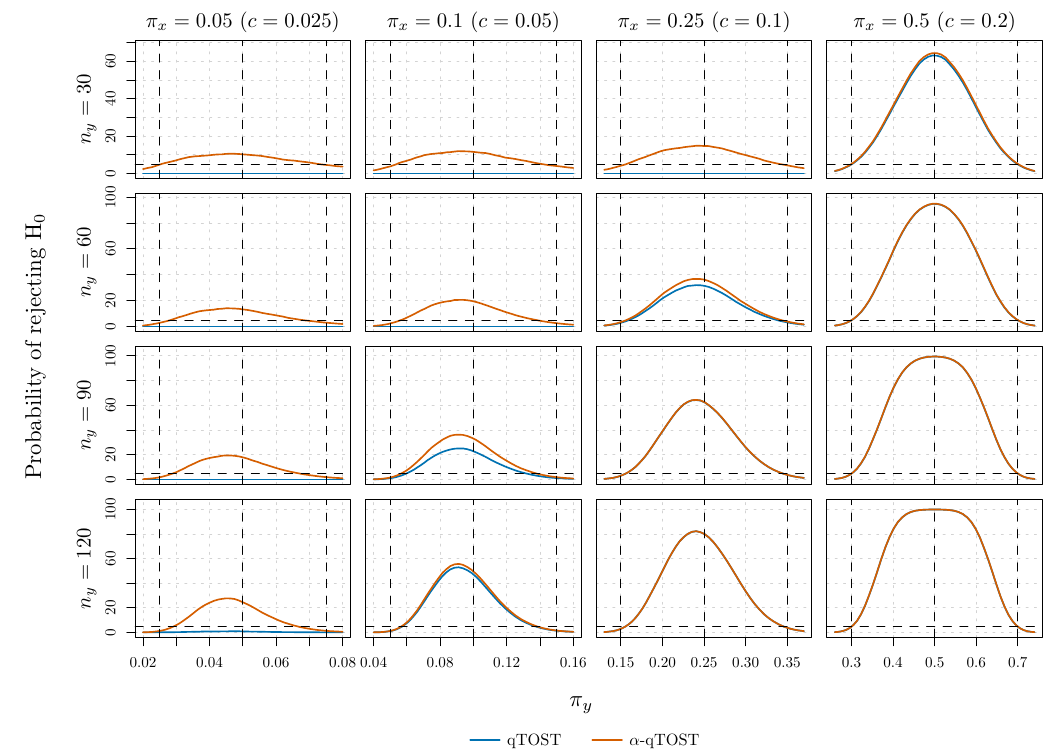}
    \caption{Simulation results comparing the probability of rejecting H$_0$ for $\alpha$-qTOST and qTOST when $\gamma=2$ and $l=\nicefrac{1}{2}$.}
    \label{fig:power_sim_setting1_gamma_2_l_2}
\end{figure}

In this section, we present additional simulation results for the study presented in Section~\ref{sec:simul_univ}.
In particular, Figures~\ref{fig:power_sim_setting1_gamma_1_l_1}-\ref{fig:power_sim_setting1_gamma_2_l_2} extend Figure~\ref{fig:sim_subset} and respectively report the cases with: (i) $\gamma=1$ and $l=1$, (ii) $\gamma=1$ and $l=\nicefrac{1}{2}$, (iii) $\gamma=1$ and $l=\nicefrac{1}{3}$, (iv) $\gamma=2$ and $l=1$, (v) $\gamma=2$ and $l=\nicefrac{1}{2}$.

\newpage
\clearpage
\subsection{Simultaneous testing of multiple quantiles}
\label{app:simul_mvt}

\begin{figure}[t]
    \centering
    \includegraphics[width=1\textwidth]{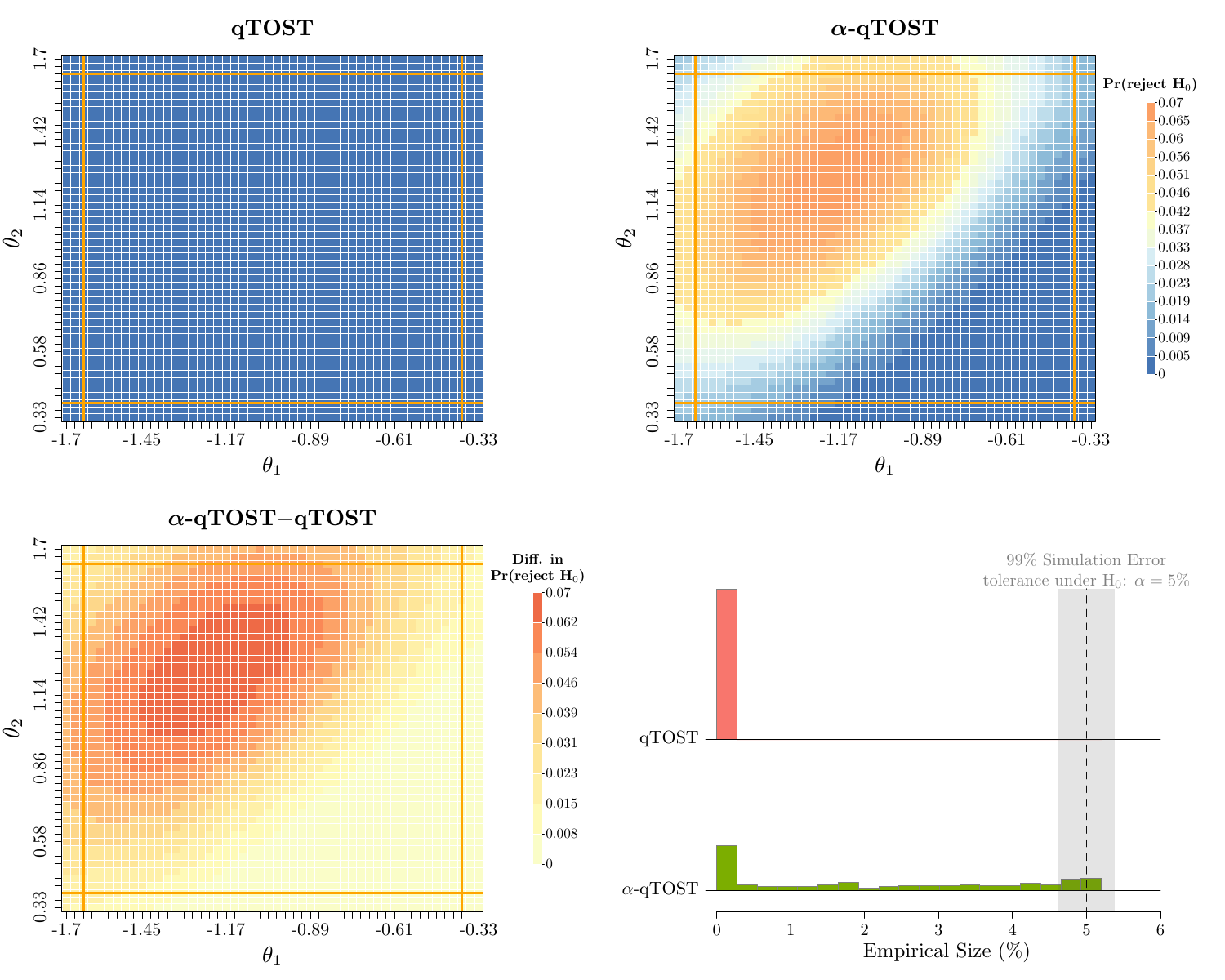}
    \caption{Simulation results comparing the operating characteristics of the qTOST and $\alpha$-qTOST procedures for $n_y = 10$.
    The heatmaps represent the probability of rejecting H$_0$ for the qTOST (top left) and $\alpha$-qTOST (top right) procedures across a grid of $\btheta$ values, as well as the difference between these probabilities (bottom left). For each method, the probability of rejecting H$_0$ along $\btheta$ values that lie on the boundary of the hypothesis space is also reported (bottom right).}
    \label{fig:sim_mvt_2}
\end{figure}

\begin{figure}[ht!]
    \centering
    \includegraphics[width=1\textwidth]{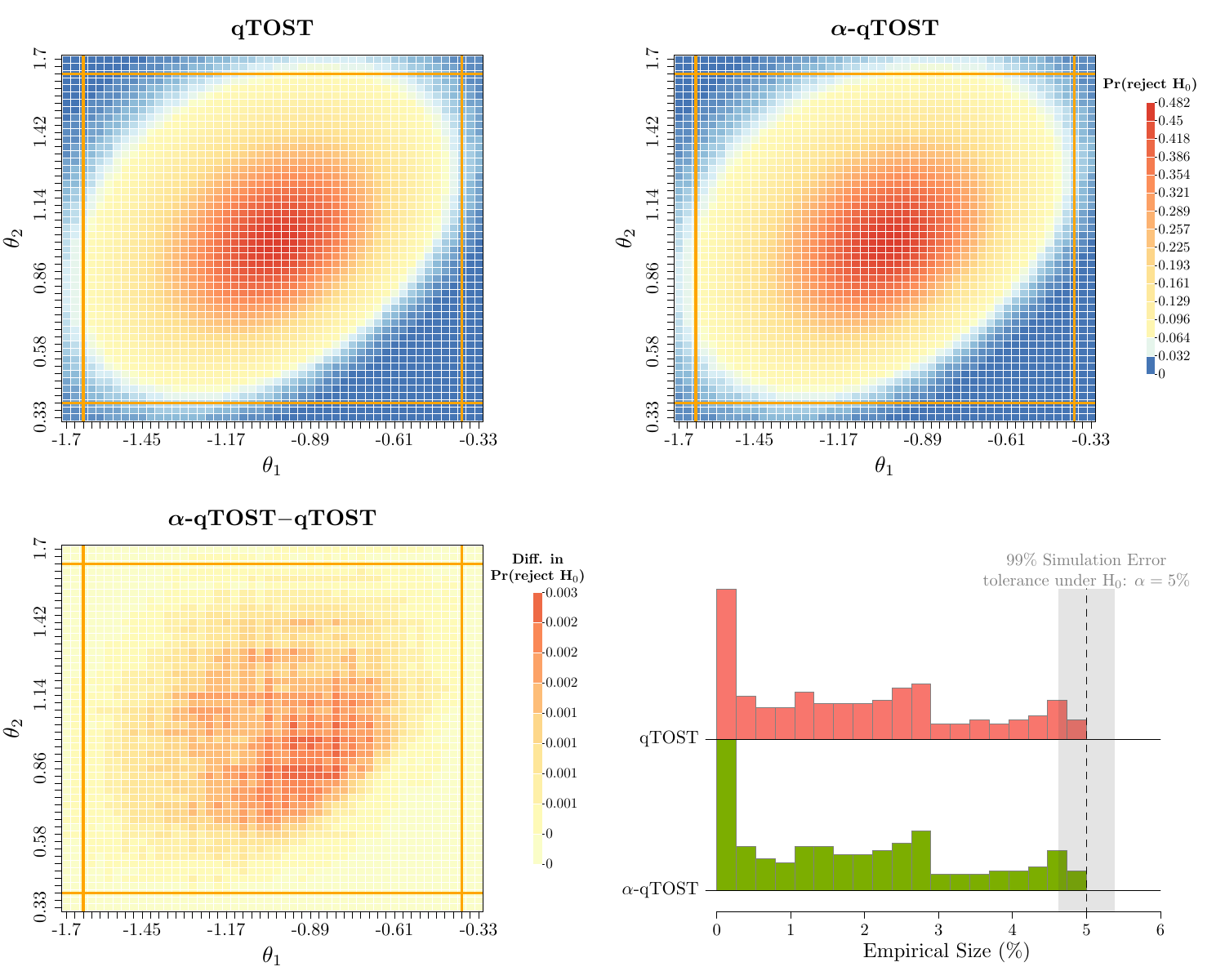}
    \caption{Simulation results comparing the operating characteristics of the qTOST and $\alpha$-qTOST procedures for $n_y = 50$.
    The heatmaps represent the probability of rejecting H$_0$ for the qTOST (top left) and $\alpha$-qTOST (top right) procedures across a grid of $\btheta$ values, as well as the difference between these probabilities (bottom left). For each method, the probability of rejecting H$_0$ along $\btheta$ values that lie on the boundary of the hypothesis space is also reported (bottom right).}
    \label{fig:sim_mvt_3}
\end{figure}

In this section, we present additional simulation results for the study presented in Section~\ref{sec:simul_mvt}.
In particular, Figures~\ref{fig:sim_mvt_2}-\ref{fig:sim_mvt_3} extend Figure~\ref{fig:sim_mvt_1} and respectively report the cases with: (i) $n_y=10$ and (ii) $n_y=50$.

\end{document}